\documentclass[journal]{IEEEtran}
\IEEEoverridecommandlockouts
\usepackage{cite}
\usepackage{graphicx}
\usepackage{tikz}
\usetikzlibrary{arrows.meta,calc}
\usepackage{epstopdf}
\usepackage{amsmath,amssymb,amsfonts}
\usepackage{algorithmic}
\usepackage{textcomp}
\usepackage{xcolor}
\usepackage[ruled,lined,linesnumbered]{algorithm2e}
\usepackage{makecell}
\usepackage{extarrows}
\usepackage{multicol}
\usepackage{multirow}
\usepackage{bm}
\usepackage{colortbl}
\usepackage{amsthm}
\usepackage{mathrsfs}
\usepackage{booktabs}
\usepackage{siunitx}
\usepackage{tabularx,booktabs}
\usepackage[flushleft]{threeparttable}
\usepackage[short]{optidef}
\usepackage{placeins}

\usepackage{xcolor}
\usepackage{xparse}
\usepackage[normalem]{ulem}

\newcommand{\defineauthor}[3]{%
  \expandafter\gdef\csname author@#1@name\endcsname{#2}%
  \expandafter\gdef\csname author@#1@color\endcsname{#3}%
}

\NewDocumentCommand{\note}{m m}{%
  \ifshownotes
    \textcolor{\csname author@#1@color\endcsname}
    {\textbf{[\csname author@#1@name\endcsname: #2]}}
  \fi
}

\NewDocumentCommand{\add}{m m}{%
  \ifshownotes
    \textcolor{\csname author@#1@color\endcsname}
    {\textbf{#2}}
  \else
    #2
  \fi
}

\NewDocumentCommand{\del}{m m}{%
  \ifshownotes
    \textcolor{\csname author@#1@color\endcsname}
    {\sout{#2}}
  \fi
}

\NewDocumentCommand{\rep}{m m m}{%
  \ifshownotes
    {\textcolor{\csname author@#1@color\endcsname}{\textbf{#3}}}%
    {\textcolor{\csname author@#1@color\endcsname}{\;{\scriptsize(was: #2)}}}%
  \else
    #3%
  \fi
}
\defineauthor{GLM5}{GLM-5}{purple}
\defineauthor{SK}{Codex}{blue}
\defineauthor{SP}{Supervisor}{red}

\definecolor{proposedblue}{RGB}{244,247,250}
\definecolor{proposedblueB}{RGB}{235,239,245}

\newcolumntype{Y}{>{\centering\arraybackslash}X}
\newcommand{\proposedrow}{\rowcolor{proposedblue}}

\newcommand{\cvprwidetablestyle}{%
  \scriptsize
  \setlength{\tabcolsep}{2.6pt}%
  \renewcommand{\arraystretch}{1.08}%
  \setlength{\aboverulesep}{0.42ex}%
  \setlength{\belowrulesep}{0.42ex}%
}
\newcommand{\cvprnarrowtablestyle}{%
  \scriptsize
  \setlength{\tabcolsep}{2.2pt}%
  \renewcommand{\arraystretch}{1.08}%
  \setlength{\aboverulesep}{0.42ex}%
  \setlength{\belowrulesep}{0.42ex}%
}

\usepackage{booktabs}

\usepackage{xcolor}

\usepackage{pifont}

\usepackage{array}
\usepackage{url}

\definecolor{refblue}{RGB}{0,80,160}
\definecolor{linkpink}{RGB}{210,45,125}
\usepackage[
  colorlinks=true,
  linkcolor=linkpink,
  citecolor=linkpink,
  urlcolor=linkpink,
  filecolor=linkpink,
  bookmarks=false
]{hyperref}

\newcommand{\cmark}{\textcolor{green!60!black}{\ding{51}}}

\newcommand{\xmark}{\textcolor{gray}{--}}

\def\BibTeX{{\rm B\kern-.05em{\sc i\kern-.025em b}\kern-.08em
T\kern-.1667em\lower.7ex\hbox{E}\kern-.125emX}}

\begin{document}

\title{Model-Driven Deep Learning with Rank-One Sensing for Efficient CSI Feedback}

\author{
Shunpu Tang,
Qianqian Yang,
Seung-Woo Ko,
Jihong Park,
and Kaibin Huang
\thanks{This work is partly supported by the National Key R\&D Program of China under Grant No. 2024YFE0200802, partly by NSFC under grant No.62293481/ 62201505/625B2165, and partly supported by the China Scholarship Council (No. 202406320381).}
\thanks{S. Tang and Q. Yang are with the College of Information Science and Electronic Engineering, Zhejiang University, Hangzhou, China (e-mail: \{tangshunpu, qianqianyang20\}@zju.edu.cn).}%
\thanks{S. Ko is with the Department of Smart Mobility Engineering, Inha University, Incheon, Korea (e-mail: swko@inha.ac.kr).}%
\thanks{J. Park is with the Information Systems Technology and Design Pillar, Singapore University of Technology and Design, Singapore (e-mail: jihong\_park@sutd.edu.sg).}
 \thanks{K. Huang is with the Department of Electrical and Electronic Engineering,
The University of Hong Kong, Hong Kong (e-mail: huangkb@hku.hk).}
\thanks{The corresponding author is Q. Yang.}
}

\maketitle
\begin{abstract}
Downlink channel state information (CSI) feedback is essential for beamforming optimization in frequency-division duplex (FDD) massive multiple-input multiple-output (MIMO) systems. However, the feedback overhead increases with the number of antennas and subcarriers, posing a major challenge to practical deployment. Although recent deep learning (DL)-based methods reduce this overhead by compressing CSI at the user equipment (UE) and reconstructing it at the base station (BS), most of them treat CSI as a generic image and rely on convolutional or Transformer architectures. As a result, the low-rank multipath prior is insufficiently exploited, while non-negligible computation is introduced at both the UE and the BS. To address this issue, we propose a low-rank prior (LRP)-guided  CSI feedback framework that performs structured sensing and reconstruction directly over rank-one channel components. Specifically, the LRP encoder conducts learnable rank-one sensing to obtain path-aware measurements at the UE, while the LRP decoder reconstructs CSI through rank-one synthesis at the BS, avoiding the iterative recovery required by classical solvers. To compensate for finite-rank approximation residuals, we extend our previous DCRNet into DCRNetV2 by incorporating the proposed LRP backbone and gated dilated-convolutional residual branches. Experiments on multiple datasets and scenarios show that the proposed methods achieve a better accuracy-complexity tradeoff than existing model-based and DL-based baselines. In particular, standalone LRP achieves more than $10$~dB NMSE improvement over model-based baselines with much lower complexity, while DCRNetV2 achieves comparable accuracy to state-of-the-art Transformer-based methods with only about one-fifth of the complexity. The source code is available at \url{https://github.com/tangshunpu/DCRNetV2}.
\end{abstract}


\section{Introduction}

\subsection{Background}
Massive multiple-input multiple-output (MIMO) is a key technology for next-generation wireless networks, as it can substantially improve spectral and energy efficiency~\cite{larsson2014massive}. By equipping the base station (BS) with many antennas, massive MIMO enables spatial multiplexing and directional beamforming to serve multiple users simultaneously. To fully exploit these gains, the BS requires accurate instantaneous downlink channel state information (CSI) for beamforming design~\cite{love2008overview}. In time-division duplex (TDD) systems, the BS can obtain downlink CSI from uplink pilots via channel reciprocity. In frequency-division duplex (FDD) systems, however, such reciprocity generally does not hold, and thus the user equipment (UE) must estimate the downlink CSI and feed it back through a limited uplink control channel~\cite{sim2016compressed}. In wideband orthogonal frequency-division multiplexing (OFDM) systems with large antenna arrays and many subcarriers, the CSI matrix becomes high-dimensional, making direct feedback impractical. 

Early studies applied compressed sensing to reduce the feedback overhead by exploiting channel sparsity~\cite{gao2018compressive}. However, these methods rely on sparsity assumptions that are sensitive to model mismatch, and their iterative inverse-problem solvers incur high BS-side latency and energy consumption~\cite{qin2018sparse}. Recently, deep learning (DL)-based methods have learned CSI compression and reconstruction in an end-to-end manner without explicit sparsity assumptions~\cite{CsiNet,guo2022overview}. Most of them treat CSI as a generic image and use convolutional or Transformer-based autoencoders, achieving higher reconstruction accuracy than conventional model-based methods, especially at high compression ratios~\cite{CsiNetPlus,TransNet}. However, their considerable computation at both the UE and the BS limits their use in resource-constrained devices and massive machine-type communications (mMTC) scenarios. Although lightweight networks~\cite{DCRNet,CLNet} reduce this cost, they often sacrifice reconstruction accuracy.

To address this issue, recent works have explored model-driven DL, which incorporates  channel-structure priors into network design~\cite{cao2023deepunfolding,he2019modeldriven}. These priors reduce the need for over-parameterized networks and improve interpretability. However, many model-driven methods still rely on NN-based iterative solvers at the BS, keeping the overall complexity high. More importantly, the low-rank multipath prior of CSI has not been fully exploited~\cite{Joint_Sparse_and_Low_Rank,Low_rank_and_Sparse_Channel_Estimation}. Under standard far-field geometric massive MIMO-OFDM models, the angular-delay CSI is dominated by a small number of separable delay-angular path responses, each of which contributes a rank-one component. Thus, the channel is naturally low-rank or approximately low-rank~\cite{Vlachos2019Wideband,zhou2017lowranktensor,shao2022exploiting}, providing a principled way to reduce both UE- and BS-side complexity. This motivates a CSI feedback architecture in which the low-rank prior (LRP) serves as the main mechanism for user-side compression and BS-side reconstruction, while lightweight neural refinement is used only to compensate for components not captured by the low-rank model.

\subsection{Related Works}
\subsubsection{Model-based CSI feedback}

Early CSI feedback designs in FDD massive MIMO followed the compressed-sensing paradigm. The user side applies a random projection to the vectorized angular-delay channel, and the BS recovers the channel by solving a sparse-recovery problem \cite{rao2014distributed}. Specifically, orthogonal matching pursuit (OMP)~\cite{omp_ref} can be used to reconstruct the CSI by iteratively selecting the most correlated sensing vector,  estimating its strength, and removing its effect until the main sparse channel components are recovered. Iterative shrinkage-thresholding algorithm (ISTA) and its accelerated variant fast-ISTA (FISTA)~\cite{fista} solve a relaxed $\ell_1$-regularized least-squares problem through iterative shrinkage. Although these methods are interpretable, their performance degrades under model mismatch, where the sparsity assumption does not always hold, and their complexity grows quickly with the antenna and subcarrier dimensions, which limits their applicability to large-scale systems.

\subsubsection{DL-based CSI feedback}
To overcome the limitations of model-based methods, CsiNet~\cite{CsiNet} pioneered DL-based CSI feedback by compressing the angular-delay-domain CSI with a convolutional autoencoder and reconstructing it at the BS, achieving clear gains over classical compressed sensing. Following this paradigm, subsequent works improved the accuracy–complexity tradeoff from different perspectives. Specifically, the authors in \cite{Cao2021Lightweight} reduced user-side complexity using channel-shuffle-based lightweight convolutions. CRNet~\cite{CRNet} extracted multi-resolution features with different receptive fields, ACRNet~\cite{ACRNet} combined aggregated residual convolutions with a width-scaling rule and DCRNet~\cite{DCRNet} used dilated convolutions to enlarge the receptive field with limited extra parameters. In addition, the authors in \cite{Sun2021Lightweight,CLNet} exploited the correlation between real and imaginary CSI components to reduce redundant computation. However, these lightweight designs usually have limited feature extraction capability, which restricts their reconstruction performance.

To pursue higher reconstruction accuracy, CsiNet+\cite{CsiNetPlus} enlarged the convolutional kernels and deepened the autoencoder to strengthen nonlinear feature extraction. Following this idea of increasing model capacity, Transformer-based methods\cite{Cheng2024SwinCFNet,TransNet} were introduced to capture long-range dependencies in CSI, and TCLNet~\cite{TCLNet} further combined convolutional blocks with Transformer modules to exploit both local structures and global interactions. More recently, large-model-based CSI feedback has also been explored. CSI-LLM~\cite{LLMCsiNet,Cui2025LLM,Guo2026PromptEnabled} reformulates CSI feedback as a sequence-to-sequence learning problem and uses an LLM to map the CSI representation into a compact feedback codeword. The authors in \cite{WiFoCF, Guo2026PromptEnabled} further extend this direction by introducing wireless foundation models, where a large model is pre-trained on diverse channel datasets and then adapted to different out-of-distribution scenarios, demonstrating strong generalization capability. Although these methods achieve improved reconstruction accuracy and robustness, they usually rely on deeper networks, attention modules, or large pre-trained models, leading to significantly higher computational cost.

\subsubsection{Model-driven DL for CSI feedback}

To combine the interpretability of model-based methods and further improve the accuracy of DL-based methods, model-driven DL~\cite{he2019modeldriven} has been developed for CSI feedback. Specifically, L2O methods~\cite{ms4l2o} unfold classical iterative solvers into trainable networks, thereby preserving an interpretable recovery process. Based on this idea, CSI-L2O~\cite{Ma2025LowComplexity} integrates compressive sensing with L2O, using a learnable linear projection at the UE for low-complexity compression and an NN-based iterative solver at the BS for CSI reconstruction. Similarly, FISTA-Net unfolds FISTA with adaptive thresholding for sparse CSI recovery~\cite{Guo2022ModelDriven}, while MarkovNet and DIFNet exploit temporal correlation through differential feedback and unfolding designs~\cite{9513579,Zhang2023ModelDriven}. Moreover, IdasNet removes redundant CSI patches through a self-information model before compression and recovery~\cite{9768327}. Although these methods improve CSI feedback efficiency by incorporating channel priors and optimization structures, their BS-side iterative reconstruction still introduces non-negligible computational cost~\cite{Chen2025CSICompression}.

\subsubsection{Low-rank perspective}
The low-rank property of CSI has been widely recognized in channel acquisition and estimation because wideband massive MIMO channels are usually generated by a small number of scattering clusters, where each path contributes a separable delay-angular response~\cite{Vlachos2019Wideband,Joint_Sparse_and_Low_Rank,zhou2017lowranktensor,shao2022exploiting}. Existing works exploit this property from different perspectives. The authors in \cite{Shen2015Joint} formulated joint multi-user CSI acquisition as a low-rank matrix completion problem by using the correlation among scheduled users. The authors in \cite{Joint_Sparse_and_Low_Rank, zhou2017lowranktensor} used low rank together with sparsity or tensor decomposition for channel estimation, while the work in ~\cite{wei2021randomized} compressed CSI by feeding back sampled submatrices and reconstructing the full channel through randomized low-rank approximation. More recently, Guo et al.~\cite{Guo2022ModelDriven} proposed a two-stage low-rank feedback scheme that decomposes CSI into a basis part and a linearly represented residual part, followed by unfolded FISTA-based reconstruction. These studies demonstrate the value of the low-rank prior, but they mainly use it as a completion constraint, a submatrix approximation, or a reconstruction-side decomposition. Moreover, the sensing and reconstruction processes are not fully co-designed according to the path-wise low-rank structure, and still rely on iterative solvers at the BS.
% In contrast, our goal is to make the path-wise rank-one structure the main information carrier of CSI feedback: the UE obtains measurements through learnable rank-one sensing matched to delay-angular components, and the BS reconstructs CSI by rank-one synthesis, with neural modules used only to refine the finite-rank and off-grid residuals.

\subsection{Contributions}
In this paper, we revisit CSI feedback from a low-rank perspective and propose a model-driven DL framework that explicitly embeds the low-rank multipath prior into both the encoder and the decoder. Instead of treating CSI feedback as a generic image compression task, we formulate it as a structured sensing and reconstruction problem over rank-one channel components. At the user side, the proposed LRP module performs learnable rank-one matched filtering over the angular-delay CSI, producing path-aware measurements with much lower complexity than convolutional encoding. At the BS side, LRP reconstructs the CSI through rank-one synthesis from the feedback codeword, which keeps the main reconstruction path interpretable and avoids the explicit reconstruction of the full CSI matrix  equired by classical solvers. To address finite-rank approximation errors and local residual structures beyond the rank-one prior, we further extend our previous DCRNet into DCRNetV2, which attaches gated dilated-convolutional residual branches to the LRP backbone while preserving the rank-one path as the dominant representation. The main contributions are summarized as follows.

\begin{itemize}
    \item We formulate CSI feedback as structured sensing and reconstruction over rank-one channel components and propose LRP. At the UE side, LRP performs learnable rank-one sensing with paired delay- and angular-domain sensing vectors, while at the BS side it reconstructs CSI through low-rank synthesis from rank-one factors. This design makes the dominant multipath prior explicit, jointly optimizes sensing and reconstruction, and avoids the iterative full-matrix recovery used by classical solvers.

    \item We design DCRNetV2 by attaching dilated-convolutional residual branches to the LRP backbone without replacing the rank-one main path. The DCR encoder, factored residual decoder, and final refinement module compensate for finite-rank approximation errors, off-grid leakage, and local cluster residuals, enabling the model to scale from lightweight to high-accuracy variants while preserving the physical interpretability of LRP.
    \item  We introduce Sionna-RT-Mix5, a new ray-tracing benchmark for CSI feedback that covers five outdoor urban and campus scenes with diverse LoS and NLoS propagation conditions. We further include a held-out Shenzhen CBD scenario to evaluate out-of-distribution generalization and adaptation with limited fine-tuning data.

\item We conduct extensive experiments on COST 2100, Sionna-RT-Mix5, and real FDD massive MIMO data to compare the proposed methods with representative model-based and DL-based baselines. We also measure on-device inference latency and energy consumption to evaluate their practical efficiency. The results show that standalone LRP achieves more than $10$~dB NMSE improvement over model-based baselines with much lower complexity, while DCRNetV2 achieves comparable accuracy to state-of-the-art Transformer-based methods with only about one-fifth of the FLOPs.
\end{itemize}

\subsection{Organization}
{\color{black}
The rest of the paper is organized as follows. Section~\ref{sec:system_model} formulates the FDD massive MIMO-OFDM CSI feedback problem. Sections~\ref{sec:proposed_method} and~\ref{sec:dcrnetv2} detail the proposed LRP framework and DCRNetV2 architecture, respectively. Section~\ref{sec:simulation} evaluates the performance of the proposed methods through simulations. Section~\ref{sec:conclusion} concludes this paper.

Notations: Bold lowercase $\bm{x}$ and uppercase $\bm{X}$ denote vectors and matrices; $(\cdot)^T$, $(\cdot)^*$, and $(\cdot)^H$ denote transpose, conjugate, and Hermitian transpose. $\Vert{}\cdot\Vert{}_F$, $\mathbb{E}[\cdot]$, and $\mathcal{CN}(\mu, \sigma^2)$ represent the Frobenius norm, expectation, and circularly symmetric complex Gaussian distribution. $\mathbb{C}^{M\times N}$ is the space of $M\times N$ complex matrices.
}
\section{System Model}\label{sec:system_model}

We consider an FDD massive MIMO-OFDM system, where the BS is equipped with \textcolor{black}{$N_t \gg 1$ transmit antennas} and the UE has a single antenna over $N_c$ subcarriers. The downlink received signal at the $n$-th subcarrier can be expressed as
\begin{align}\label{receive signal}
    y_n=\bm{h}_{n}^H\bm{b}_n x_n+z_n,
\end{align}
where $\bm{h}_n \in \mathbb{C}^{N_t \times 1}$ denotes the complex-valued CSI vector for the $n$-th subcarrier, $\bm{h}_n^H$ denotes its conjugate transpose, $\bm{b}_n \in \mathbb{C}^{N_t \times 1}$ is the beamforming vector, $x_n$ is the transmitted symbol, and $z_n \sim \mathcal{CN}(0,\sigma^2)$ is the complex additive noise. The CSI across all subcarriers can be collected as $\bm{H}=[\bm{h}_1,\bm{h}_2,\cdots,\bm{h}_{N_c}]^T \in \mathbb{C}^{N_c\times N_t}$, which contains $2N_c N_t$ real-valued coefficients.

To obtain a better quality of service, the UE needs to feed back the downlink CSI $\bm{H}$ to the BS, which then uses it to optimize the beamforming vectors $\{\bm{b}_n\}$. However, directly feeding back $\bm{H}$ would require $2N_c N_t$ real-valued feedback symbols, which is impractical for large $N_c$ and $N_t$ given the limited bandwidth of the uplink control channel. Therefore, the CSI must be compressed at the UE before feedback. Specifically, the CSI matrix $\bm{H}$ can be transformed into the angular-delay domain by a 2D-discrete Fourier transform (2D-DFT), in which the channel exhibits a sparse structure, given by
\begin{align}\label{DFT}
    \bm{\tilde{H}}=\bm{F}_c \bm{H} \bm{F}_t^{H} ,
\end{align}
where $\bm{F}_c\in\mathbb{C}^{N_c\times N_c}$ and $\bm{F}_t\in\mathbb{C}^{N_t\times N_t}$ transform the subcarrier and antenna dimensions into the delay and angular domains, respectively. \textcolor{black}{After the 2D-DFT, most of the energy of $\tilde{\bm{H}}$ is concentrated in the first $N_a\leq N_c$ delay-domain rows.} The remaining $N_c-N_a$ rows contain near-zero values, and we therefore truncate $\bm{\tilde{H}}$ to its first $N_a$ rows and denote the resulting angular-delay CSI as $\bm{H}_a \in \mathbb{C}^{N_a\times N_t}$. To further reduce the feedback overhead, the UE applies an encoder $\mathcal{E}_{\theta_1}(\cdot)$ to compress $\bm{H}_a$ into a low-dimensional codeword, given by
\begin{equation}
    \bm v
    =
    \mathcal{E}_{\theta_1}(\bm H_a),
    \quad
    \bm v\in\mathbb{R}^{M},
    \label{eq:system_encoder}
\end{equation}
\textcolor{black}{where $M\in\mathbb{Z}^{+}$ denotes the number of real-valued feedback symbols. The corresponding compression ratio is defined as
$\eta={M}/{2N_aN_t}$.}

Upon receiving the codeword $\bm v$, the BS reconstructs the angular-delay CSI through a decoder $\mathcal{D}_{\theta_2}(\cdot)$, given by
\begin{equation}
    \hat{\bm H}_a
    =
    \mathcal{D}_{\theta_2}(\bm v).
    \label{eq:system_decoder}
\end{equation}
The objective of the encoder-decoder pair is to minimize the reconstruction error between the original and the reconstructed CSI, which is measured by the mean squared error (MSE), given by
\begin{equation}
    \label{object}
    \min_{\theta_1,\theta_2}
    \mathbb{E}\!\left[
    \left\|\bm H_a-\hat{\bm H}_a\right\|_F^2
    \right].
\end{equation}

\section{Low-Rank Prior-based CSI Compression}\label{sec:proposed_method}

\begin{figure}[t]
    \centering
    \includegraphics[width=0.95\columnwidth]{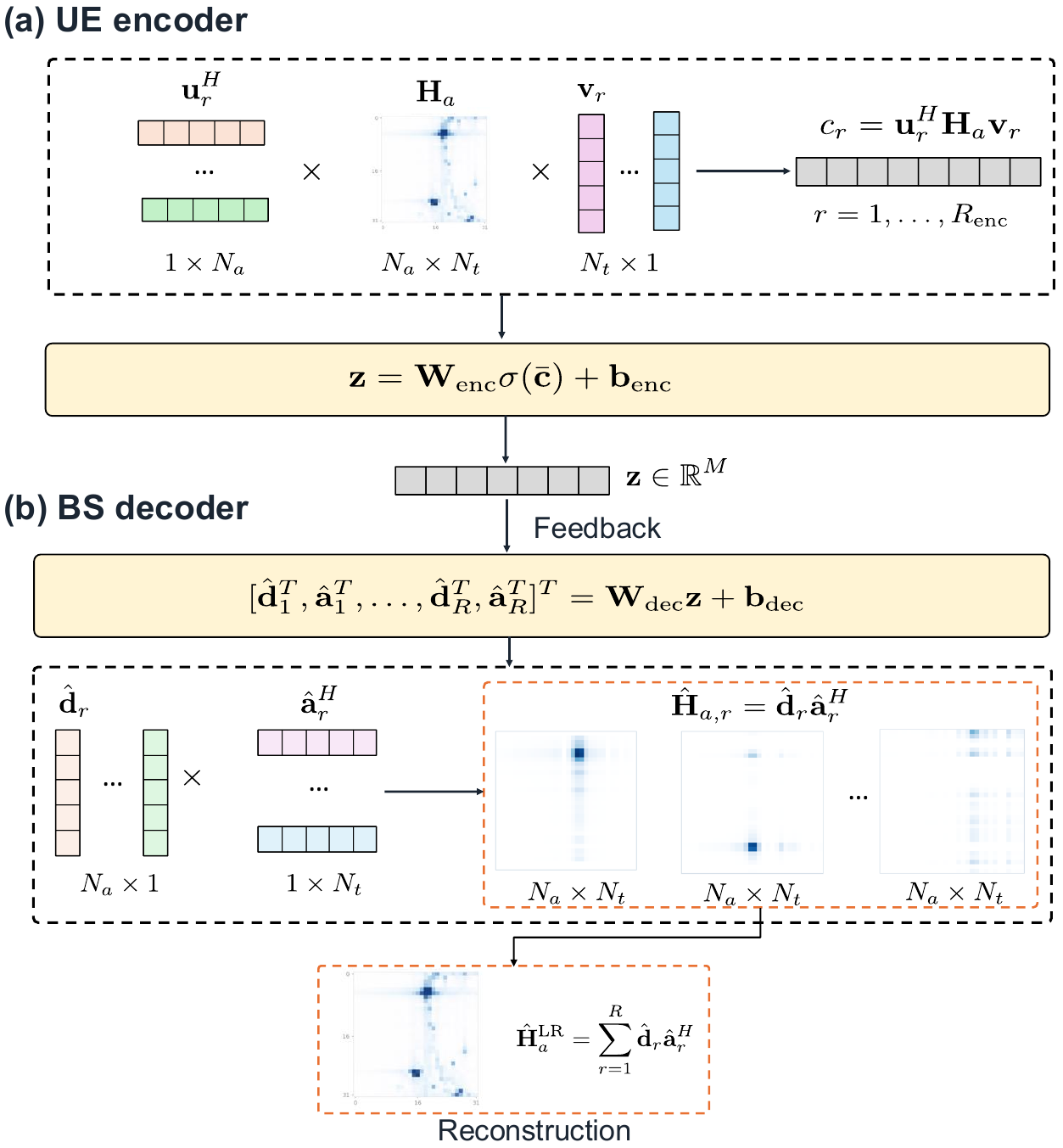}
    \caption{LRP CSI feedback: (a) rank-one sensing and compression; (b) factor generation and rank-one synthesis. Cell shading is illustrative.}
    \label{fig:lrp_arch}
\end{figure}

In this section, we recast CSI feedback as a structured sensing and reconstruction problem and present \textcolor{black}{the proposed LRP framework as shown in \autoref{fig:lrp_arch}}, where the \textcolor{black}{LRP encoder} performs rank-one sensing to obtain a low-dimensional measurement and the \textcolor{black}{LRP decoder} reconstructs the CSI from the measurement.  We first introduce the low-rank multipath prior of the angular-delay channel, and then present the detailed design of the encoder and decoder. Finally, we describe the optimization procedure and analyze the computational complexity.

\subsection{Low-Rank Multipath Prior}

After the 2D-DFT in \eqref{DFT}, the truncated angular-delay CSI matrix
\(\bm H_a\in\mathbb{C}^{N_a\times N_t}\) is not a generic dense matrix, but is governed by the underlying multipath propagation structure. We now derive this prior from the frequency-spatial channel model. Under a standard far-field geometric channel with a ULA at the BS, we assume that the path gains are frequency independent over the considered OFDM bandwidth, the array response is frequency invariant, and the delay and spatial responses are separable. Then, the \(n\)-th subcarrier channel row can be written as \cite{Vlachos2019Wideband, Joint_Sparse_and_Low_Rank, zhou2017lowranktensor}
\begin{align}
    \bm H_{n,:}
    =
    \sum_{\ell=1}^{L}
    \alpha_\ell e^{-j2\pi f_n\tau_\ell}
    \bm a^{H}(\theta_\ell),
    \quad n=1,\ldots,N_c,
    \label{eq:freq_spatial_geo_channel}
\end{align}
where \(\alpha_\ell\), \(\tau_\ell\), and \(\theta_\ell\) denote the complex path gain, delay, and angle of the \(\ell\)-th path, respectively, \(f_n\) is the baseband frequency of the \(n\)-th subcarrier, and \(\bm a(\theta_\ell)\in\mathbb{C}^{N_t}\) is the array steering vector. By stacking all subcarriers, \eqref{eq:freq_spatial_geo_channel} can be rewritten as
\begin{align}
    \bm H
    =
    \sum_{\ell=1}^{L}
    \alpha_\ell
    \bm q(\tau_\ell)\bm a^{H}(\theta_\ell),
    \label{eq:freq_spatial_stacked_channel}
\end{align}
where
\(\bm q(\tau_\ell)=[e^{-j2\pi f_1\tau_\ell},\ldots,e^{-j2\pi f_{N_c}\tau_\ell}]^{T}\in\mathbb{C}^{N_c}\)
is the frequency-domain delay response. Let \(\bm S_a=[\bm I_{N_a},\bm 0]\in\mathbb{R}^{N_a\times N_c}\) denote the row-selection matrix that keeps the first \(N_a\) delay bins. Combining \eqref{DFT} with \eqref{eq:freq_spatial_stacked_channel}, the truncated angular-delay CSI is obtained as
\begin{align}
    \bm H_a
    &=
    \bm S_a\bm F_c\bm H\bm F_t^{H} \notag\\
    &=
    \sum_{\ell=1}^{L}
    \alpha_\ell
    \big(\bm S_a\bm F_c\bm q(\tau_\ell)\big)
    \big(\bm F_t\bm a(\theta_\ell)\big)^{H}.
    \label{eq:geo_low_rank}
\end{align}
This expression shows that each physical path remains an outer product of a transformed delay response and a transformed angular response after the 2D-DFT and delay-row truncation. Therefore, the multipath channel is represented as a sum of rank-one delay-angular components. Since premultiplication and postmultiplication by DFT matrices, together with delay-row selection, cannot increase the matrix rank, \(\bm H_a\) satisfies
\begin{align}
    \operatorname{rank}(\bm H_a)
    &\leq
    \sum_{\ell=1}^{L}
    \operatorname{rank}\!\left(
    \alpha_\ell
    \big(\bm S_a\bm F_c\bm q(\tau_\ell)\big)
    \big(\bm F_t\bm a(\theta_\ell)\big)^{H}
    \right) \notag\\
    &\leq L.
    \label{eq:rank_bound}
\end{align}
When the number of dominant paths is much smaller than the matrix dimensions, i.e.,
\(L\ll\min(N_a,N_t)\), \(\bm H_a\) is therefore low-rank or can be accurately approximated by a low-rank matrix \cite{Low_rank_and_Sparse_Channel_Estimation, shao2022exploiting}.  Motivated by this observation, we adopt the low-rank structure as the working prior and approximate \(\bm H_a\) by \(R\) rank-one components, given by
\begin{align}
    \bm H_a
    \approx
    \sum_{r=1}^{R}\bm d_r\bm a_r^{H},
    \label{eq:low_rank_prior}
\end{align}
where \(R\) is a designable model rank that controls the number of rank-one
components used by the architecture and need not equal the physical path number
\(L\).

\subsection{Rank-One Sensing-Based CSI Compression}

The low-rank prior in \eqref{eq:low_rank_prior} suggests that \(\bm H_a\) is not an arbitrary random matrix and cannot simply be regarded as a generic image. Instead, it has an explicit physical structure that is generated by a small number of multipath components, and each component is rank-one and separable across the delay and angular domains. This observation motivates the use of rank-one sensing, under which a low-rank matrix can be efficiently recovered from a small number of rank-one measurements formed by pairs of sensing vectors along the row and column dimensions \cite{cai2015rop, zhong2015efficient,pmlr-v89-li19e,rank_one_prior}. By designing the sensing vectors to align with the delay and angular steering vectors, we can obtain measurements that are directly informative about the underlying multipath structure, rather than relying on generic convolutional filters to learn this structure implicitly.

Specifically,  for the \(r\)-th sensing branch, we construct the measurement operator as a rank-one matrix
\(\bm A_r=\bm u_r\bm v_r^{H}\in\mathbb{C}^{N_a\times N_t}\) from two complex-valued
sensing vectors \(\bm u_r\in\mathbb{C}^{N_a}\) and
\(\bm v_r\in\mathbb{C}^{N_t}\), which yields the bilinear measurement, given by
\begin{align}
    c_r
    =
    \left\langle \bm A_r, \bm H_a \right\rangle_F
    =
    \bm u_r^{H}\bm H_a \bm v_r,
    \qquad r=1,\ldots,R_{\mathrm{enc}},
    \label{eq:lpa_rank_one_sensing}
\end{align}
where \(\langle\cdot,\cdot\rangle_F\) denotes the complex Frobenius inner
product. Notably, substituting the low-rank prior \eqref{eq:low_rank_prior}
into \eqref{eq:lpa_rank_one_sensing} yields
\(c_r\!\approx\!\sum_{k=1}^{R}(\bm u_r^{H}\bm d_{k})(\bm a_{k}^{H}\bm v_r)\),
which is also similar to the matched filtering that measures the alignment between $\bm H_a$ 
and the delay--angular pattern parameterized by 
$(\bm u_r,\bm v_r)$. In other words, each sensing branch detects whether the CSI contains 
a path component whose delay--angular signature aligns with the 
pattern parameterized by $(\bm u_r,\bm v_r)$.

To achieve the desired feedback dimension \(M\), we further project the measurements \(\{c_r\}\) to a low-dimensional codeword through a lightweight MLP layer. Specifically, we stack the measurements into a vector
\(\bm c=[c_1,\ldots,c_{R_{\mathrm{enc}}}]^T\in\mathbb{C}^{R_{\mathrm{enc}}}\) and then convert it into a real-valued vector by concatenating the real and imaginary parts as \(\bar{\bm c}=[\Re\{\bm c\}^{T},\Im\{\bm c\}^{T}]^{T}\in\mathbb{R}^{2R_{\mathrm{enc}}}\). The feedback codeword is then produced by \textcolor{black}{the LRP encoder}, given by
\begin{align}
    \bm z
    =
    \bm W_{\mathrm{enc}} \sigma(\bar{\bm c})+\bm b_{\mathrm{enc}},
    \label{eq:lpa_codeword_projection}
\end{align}
where \(\bm W_{\mathrm{enc}}\in\mathbb{R}^{M\times 2R_{\mathrm{enc}}}\) and
\(\bm b_{\mathrm{enc}}\in\mathbb{R}^{M}\) are learnable parameters, and \(\sigma(\cdot)\) is
an activation function. In this way, the \textcolor{black}{LRP encoder} embeds the low-rank
delay--angular prior directly into the measurement structure, rather than
relying on generic neural layers to learn this structure implicitly without introducing larger computational overhead.

\subsection{Low-Rank Multipath Reconstruction}
At the BS side, the \textcolor{black}{LRP decoder} reconstructs the CSI from the feedback codeword with the help of the same low-rank prior. Rather than directly generating all \(N_aN_t\) complex CSI coefficients using NN layers, it instead generates the rank-one factors of each path component and then synthesizes the CSI through their outer products. Specifically, when receiving the codeword \(\bm z\), the \textcolor{black}{LRP decoder} first maps the received codeword
\(\bm z\in\mathbb{R}^{M}\) to the rank-one factors of all \(R\) reconstruction
branches, given by
\begin{align}
    \big[\hat{\bm d}_1^{T},\hat{\bm a}_1^{T},\ldots,
    \hat{\bm d}_{R}^{T},\hat{\bm a}_{R}^{T}\big]^{T}
    =
    \textcolor{black}{\bm W_{\mathrm{dec}}} \bm z+\textcolor{black}{\bm b_{\mathrm{dec}}},
    \label{eq:decoder_linear_projection}
\end{align}
where \(\textcolor{black}{\bm W_{\mathrm{dec}}}\in\mathbb{C}^{R(N_a+N_t)\times M}\) and
\(\textcolor{black}{\bm b_{\mathrm{dec}}}\in\mathbb{C}^{R(N_a+N_t)}\) are learnable parameters. The vectors
\(\hat{\bm d}_r\in\mathbb{C}^{N_a}\) and \(\hat{\bm a}_r\in\mathbb{C}^{N_t}\) are
the delay-domain and angular-domain factors of the \(r\)-th branch\footnote{The complex coefficient of each rank-one component is absorbed into the magnitudes and phases of $\hat{\bm d}_r$ and $\hat{\bm a}_r$, so no separate
scalar gain is introduced.}. The branch output
is the rank-one outer product
\begin{align}
    \hat{\bm H}_{a,r}
    =
    \hat{\bm d}_r
    \hat{\bm a}_r^{H}.
    \label{eq:decoder_rank_one_component}
\end{align}
Next, \textcolor{black}{the LRP decoder} sums the outputs of all branches to obtain the low-rank reconstruction of the CSI, given by
\begin{align}
    \hat{\bm H}_{a}^{\mathrm{LR}}
    =
    \sum_{r=1}^{R}
    \hat{\bm d}_r
    \hat{\bm a}_r^{H}.
    \label{eq:decoder_low_rank_reconstruction}
\end{align}
We note that the number of decoder branches $R$ is a reconstruction rank hyperparameter rather than the exact
number of physical propagation paths. Therefore, each branch is not required to correspond to one physical path, while  the $R$ branches jointly provide a compact low-rank approximation of the dominant CSI structure.

\begin{algorithm}[t!]
  \SetAlgoLined
  \DontPrintSemicolon
  \caption{Inference Process of the Proposed Method}\label{alg:dcrnet}
  \KwIn{Angular-delay CSI matrix 
  \(\bm H_a\in\mathbb{C}^{N_a\times N_t}\); trained parameters 
  \(\bm{\Theta}_e\) and \(\bm{\Theta}_d\).}
  \KwOut{Reconstructed CSI \(\hat{\bm H}_a\).}
  
  \textbf{UE side: rank-one sensing}\;
  \For{$r=1,\ldots,R_{\mathrm{enc}}$}{
      Compute matched-filtering coefficient 
      \(c_r\leftarrow\bm u_r^{H}\bm H_a\bm v_r\)\;
  }
  Stack coefficients 
  \(\bm c\leftarrow[c_1,\ldots,c_{R_{\mathrm{enc}}}]^{T}\in\mathbb{C}^{R_{\mathrm{enc}}}\)\;
  Form real-valued representation 
  \(\bar{\bm c}\leftarrow[\Re\{\bm c\}^{T},\Im\{\bm c\}^{T}]^{T}\)\;
  Generate feedback codeword 
  \(\bm z\leftarrow\bm W_{\mathrm{enc}}\sigma(\bar{\bm c})+\bm b_{\mathrm{enc}}\)\;
  Send \(\bm z\) to the BS over the feedback link\;
  
  \textbf{BS side: low-rank synthesis}\;
  Apply linear projection to obtain rank-one factors
  \(\{\hat{\bm d}_r,\hat{\bm a}_r\}_{r=1}^{R}
  \leftarrow \textcolor{black}{\bm W_{\mathrm{dec}}}\bm z+\textcolor{black}{\bm b_{\mathrm{dec}}}\)\;
  Reconstruct
  \(\hat{\bm H}_a\leftarrow
  \sum_{r=1}^{R}
  \hat{\bm d}_r\hat{\bm a}_r^{H}\)\;
  
  \Return{\(\hat{\bm H}_a\)}\;
  \end{algorithm}

% \begin{table*}[ht!]
%   \centering
%   \caption{Simplified computational complexity comparison of different CSI feedback methods.}
%   \label{tab:simplified_complexity}
%   \begin{tabular}{lccc}
%   \toprule
%   Methods 
%   & CSI Compression Complexity 
%   & CSI Recovery Complexity
%   & Overall Complexity \\
%   \midrule
  
%   Wavelet~\cite{wavelet_ref}
%   &
%   $\mathcal{O}\!\left(N_tN_r\log(N_tN_r)\right)$
%   &
%   $\mathcal{O}\!\left(N_tN_r\log(N_tN_r)\right)$
%   &
%   $\mathcal{O}\!\left(N_tN_r\log(N_tN_r)\right)$
%   \\
  
%   OMP-DCT~\cite{omp_ref}
%   &
%   $\mathcal{O}\!\left(N_r^2N_t^2\right)$
%   &
%   $\mathcal{O}\!\left(N_r^2N_t^2T^2\right)$
%   &
%   $\mathcal{O}\!\left(N_r^2N_t^2T^2\right)$
%   \\
  
%   OMP-KLT~\cite{omp_ref}
%   &
%   $\mathcal{O}\!\left(N_r^3N_t^3\right)$
%   &
%   $\mathcal{O}\!\left(N_r^2N_t^2T^2\right)$
%   &
%   $\mathcal{O}\!\left(N_r^3N_t^3+N_r^2N_t^2T^2\right)$
%   \\
  
%   Randomized LRA~\cite{randomized_lra_ref}
%   &
%   $\mathcal{O}\!\left(N_tN_r\right)$
%   &
%   $\mathcal{O}\!\left(rN_rN_t\right)$
%   &
%   $\mathcal{O}\!\left(rN_rN_t\right)$
%   \\
%  Ours
%   &
%   $\mathcal{O}\!\left(N_rN_t\right)$
%   &
%   $\mathcal{O}\!\left(N_rN_t(N_r+N_t)\right)$
%   &
%   $\mathcal{O}\!\left(N_rN_t(N_r+N_t)\right)$
%   \\
%   \bottomrule
%   \end{tabular}
%   \end{table*}

\subsection{Optimization}
The above design has specified the structural form of the rank-one sensing \textcolor{black}{LRP encoder} in \eqref{eq:lpa_rank_one_sensing}--\eqref{eq:lpa_codeword_projection} and the low-rank synthesis \textcolor{black}{LRP decoder} in \eqref{eq:decoder_linear_projection}--\eqref{eq:decoder_low_rank_reconstruction}, which can be jointly characterized by the structured parameter set
\begin{align}
    \bm{\Theta}
    =
    \{\mathcal{U},\mathcal{V},\bm W_{\mathrm{enc}},\bm b_{\mathrm{enc}},\textcolor{black}{\bm W_{\mathrm{dec}}},\textcolor{black}{\bm b_{\mathrm{dec}}}\},
    \label{eq:optimized_parameters}
\end{align}
where \(\mathcal{U}=\{\bm u_r\}_{r=1}^{R_{\mathrm{enc}}}\) and \(\mathcal{V}=\{\bm v_r\}_{r=1}^{R_{\mathrm{enc}}}\) are the rank-one sensing vectors. The remaining question is how to determine \(\bm{\Theta}\). A straightforward solution is to follow classical low-rank matrix sensing. \textcolor{black}{In this approach, the sensing matrices \(\{\bm A_m\}_{m=1}^{\tilde M}\) are usually generated at random, for example from a Gaussian distribution, and then kept fixed to form a linear sensing operator \(\mathcal A(\bm X)=[\langle \bm A_m,\bm X\rangle_F]_{m=1}^{\tilde M}\).} Given the measurements $\bm y=\mathcal A(\bm H_a)$, the
low-rank matrix is recovered by solving
\begin{align}
\hat{\bm H}_a
=
\arg\min_{\bm X}
\tfrac{1}{2}
\left\|
\mathcal A(\bm X)-\bm y
\right\|_2^2
+
\lambda \|\bm X\|_* ,
\label{eq:classical_nuclear_norm_recovery}
\end{align}
where $\|\bm X\|_*$ is the regularization term that promotes low-rank structure, and $\lambda$ is a regularization parameter. 

\begin{figure*}
    \centering
    \includegraphics[width=0.95\textwidth]{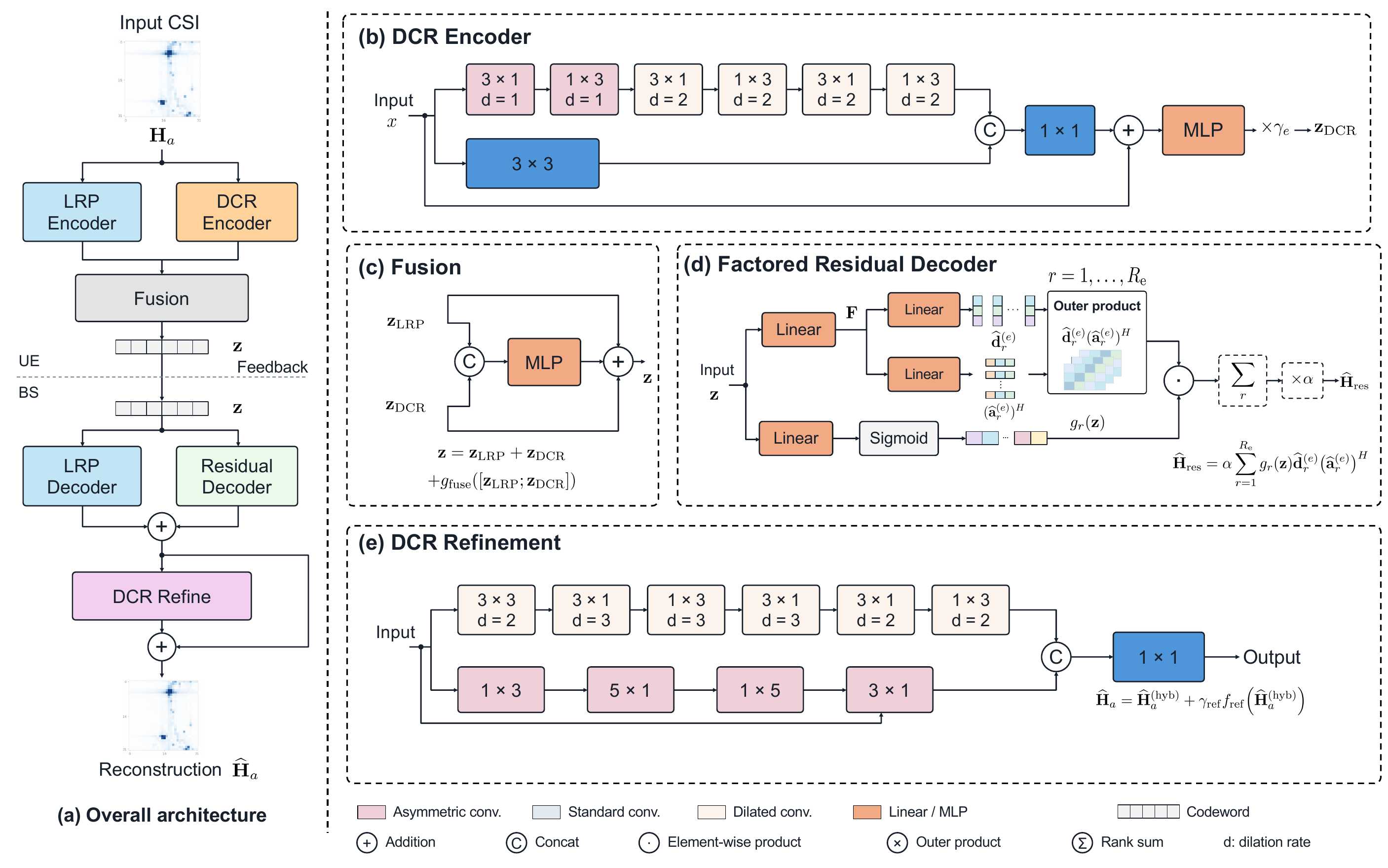}
    \caption{Architecture of DCRNetV2. The LRP backbone is shown in blue, and the convolutional residual branch is shown in orange. The two branches are fused at the encoder output, and they are decoded separately at the BS side. The final CSI estimate is obtained by summing the direct low-rank reconstruction and the factored residual reconstruction.}
    \label{fig:dcrnetv2_arch}
\end{figure*}

However, this classical formulation is not well suited to CSI feedback. First, random sensing matrices are designed to be universal for general low-rank matrices, while the angular--delay CSI matrix $\bm H_a$ follows a specific multipath structure as shown in \eqref{eq:geo_low_rank}. Therefore, random sensing patterns may not match the dominant delay--angular components
of the channel. Second, \eqref{eq:classical_nuclear_norm_recovery} requires solving an optimization problem for every received codeword, which introduces iterative online optimization at the BS. Third, the sensing operator and the recovery algorithm are designed
separately, so the obtained measurements are not necessarily optimal for the final CSI reconstruction error.

Motivated by these three points, we jointly optimize the sensing and reconstruction stages by directly minimizing the empirical reconstruction error over a training dataset. Specifically, given \(N_s\) training samples, the parameter set is obtained by minimizing the empirical reconstruction error:
\begin{align}
    \min_{\bm{\Theta}}\;
    \frac{1}{N_s}\sum_{i=1}^{N_s}
    \big\|\bm H_a^{(i)}-\hat{\bm H}_a^{(i)}\big\|_F^2,
    \label{eq:structured_parameter_optimization}
\end{align}
which is solved by mini-batch stochastic gradient-based training over the CSI samples. In this way, the rank-one sensing vectors \(\{\bm u_r,\bm v_r\}\) are adapted to the angular-delay statistics of practical CSI, and they are coupled with the synthesis decoder through the same training objective, so that each measurement \(c_r\) in \eqref{eq:lpa_rank_one_sensing} is encouraged to preserve the CSI features most useful for the final reconstruction rather than being fixed independently of the recovery stage.
\subsection{Complexity Analysis}
\label{subsec:lrp_complexity}

We now analyze the computational complexity of the proposed LRP framework. At the user side, each rank-one sensing in \eqref{eq:lpa_rank_one_sensing} is computed as \(\bm u_r^{H}(\bm H_a\bm v_r)\) at a cost of \(\mathcal{O}(N_aN_t)\), so the sensing stage with \(R_{\mathrm{enc}}\) operators scales as \(\mathcal{O}(R_{\mathrm{enc}}N_aN_t)\), and the subsequent linear projection in \eqref{eq:lpa_codeword_projection} adds \(\mathcal{O}(MR_{\mathrm{enc}})\). Therefore, the user-side complexity is \(\mathcal{O}(R_{\mathrm{enc}}(N_aN_t+M))\). At the BS side, the direct linear projection in \eqref{eq:decoder_linear_projection} costs \(\mathcal{O}(MR(N_a+N_t))\), and the rank-one synthesis in \eqref{eq:decoder_low_rank_reconstruction} sums \(R\) rank-one outer products at \(\mathcal{O}(RN_aN_t)\), giving a BS-side complexity of \(\mathcal{O}(R[M(N_a+N_t)+N_aN_t])\). The overall complexity of the LRP framework is therefore
\begin{equation}
    \mathcal{O}\!\left(R_{\mathrm{enc}}(N_aN_t+M)+R[M(N_a+N_t)+N_aN_t]\right).
    \label{eq:overall_complexity}
\end{equation}
Since the encoder uses only lightweight rank-one sensing and a linear projection, and the decoder avoids any explicit \(\mathcal{O}(N_a^{2}N_t^{2})\) matrix reconstruction, the proposed scheme incurs much lower complexity than conventional convolutional autoencoder-based feedback, and is therefore well suited for resource-constrained users.

\section{Design of DCRNetV2}\label{sec:dcrnetv2}

In this section, we present DCRNetV2, an architecture family that enhances LRP while preserving its core backbone. As shown in Fig.~\ref{fig:dcrnetv2_arch}, \textcolor{black}{the framework is designed to preserve critical channel information. At the encoder, global low-rank and local features are extracted in parallel before compression, since local details discarded by the LRP projection cannot be recovered later from the codeword. At the decoder, the codeword is first restored to the angular-delay domain via low-rank synthesis and a lightweight rank extension, followed by convolutional refinement to capture local correlations.}

\subsection{Encoder Design}
\label{subsec:dcrnetv2_combine}

At the user side, the angular-delay channel $\bm H_a$ is processed by two parallel branches. The first branch is the LRP encoder in \eqref{eq:lpa_codeword_projection}, which performs learnable rank-one sensing followed by a linear projection and produces the codeword \bm $z_{\mathrm{LRP}}\in\mathbb{R}^{M}$. \textcolor{black}{Each rank-one sensing matrix can be viewed as a learnable separable template that globally matches the angular-delay channel, making LRP effective at extracting its dominant structured components. However, pure rank-one sensing patterns struggle in rich-scattering, non-line-of-sight (NLoS) environments. In such scenarios, severe multipath delay spread and angular dispersion naturally smear channel energy across adjacent bins, forming diffuse local clusters that cannot be fully captured by a few global rank-one components alone.}

To address this issue, the second branch introduces the DCR encoder shown in Fig.~\ref{fig:dcrnetv2_arch}. It uses asymmetric $1\times k$ and $k\times 1$ convolutions to separately learn angular and delay correlations, while parallel dilated paths expand the receptive field to cover broader multipath clusters. \textcolor{black}{Moreover, the DCR branch operates directly on $\bm H_a$ in parallel with LRP because local details lost during global projection cannot be restored by subsequent post-processing. Together, the two branches extract complementary global and local features prior to compression.} Denoting the DCR operation as $f_{\mathrm{DCR}}(\cdot)$, the auxiliary codeword is given by
\begin{align}
    \bm z_{\mathrm{DCR}}
    =
    \gamma_e\,f_{\mathrm{DCR}}(\bm H_a)
    \in\mathbb{R}^{M},
    \label{eq:dcr_aux_codeword}
\end{align}
where \(\gamma_e\) is a learnable scalar gate that controls how much of the convolutional residual enters the feedback vector. Next, since the two codewords are generated by different transforms, direct element-wise addition would implicitly assume aligned feature semantics. We instead concatenate them and employ a small MLP to learn their interaction, following the widely used concatenation-based feature fusion in multi-branch networks~\cite{ronneberger2015unet}, given by
\begin{align}
    \bm z
    =
    \bm z_{\mathrm{LRP}}
    +
    \bm z_{\mathrm{DCR}}
    +
    g_{\mathrm{fuse}}\!\left(\left[\bm z_{\mathrm{LRP}};\bm z_{\mathrm{DCR}}\right]\right),
    \label{eq:dcr_lrp_encoder}
\end{align}
where \(g_{\mathrm{fuse}}(\cdot)\) is a two-layer MLP. The additive terms preserve a direct path from both branches, while the MLP learns only the residual interaction between the low-rank and convolutional codewords. To avoid training instability at the early stage, \(\gamma_e\) and the output layer of \(g_{\mathrm{fuse}}(\cdot)\) are initialized to a small value so that the initial codeword is close to \(\bm z_{\mathrm{LRP}}\) and the convolutional contribution is gradually increased when necessary.

\subsection{Decoder Design}
\label{subsec:dcrnetv2_decoder}

At the BS side, the codeword \(\bm z\) is first decoded by a low-rank synthesis module that mirrors \eqref{eq:decoder_linear_projection}--\eqref{eq:decoder_low_rank_reconstruction} and produces a direct rank-\(R\) estimate
\begin{align}
    \hat{\bm H}_a^{(\mathrm{dir})}
    =
    \sum_{r=1}^{R}\hat{\bm d}_r\hat{\bm a}_r^{H},
    \label{eq:dcrv2_direct_decoder}
\end{align}
where \(\hat{\bm d}_r\) and \(\hat{\bm a}_r\) are obtained from \(\bm z\) through the LRP linear projection. This direct path preserves the interpretable low-rank reconstruction of LRP, but its rank is hard-capped at \(R\). \textcolor{black}{When the channel exhibits rich scattering or continuous multipath dispersion, the channel energy can spread over a larger number of effective rank components, resulting in a slowly decaying rank tail that is difficult to represent with a fixed rank-$R$ model. This leads a representation bottleneck for the direct decoder. A straightforward solution is to increase $R$, but doing so linearly increases both the computational complexity of the decoder.}

To supplement the rank more efficiently, we introduce a residual decoder in Fig.~\ref{fig:dcrnetv2_arch} that generates extra rank-one components from the same codeword \(\bm z\), but with a lower cost than enlarging the direct branch. Specifically, the codeword is first mapped by the residual feature projection \(\bm W_{\mathrm{res}}\in\mathbb{R}^{R_{\mathrm{e}}d\times M}\) and reshaped into a tensor \(\bm F\in\mathbb{R}^{R_{\mathrm{e}}\times d}\), where \(d\ll M\) is a small bottleneck dimension. Each extra rank-one component is then generated from a shared factored pair, which can be expressed as
\begin{align}
    \hat{\bm d}^{(e)}_r=\bm W_d\bm F_r,\quad
    \hat{\bm a}^{(e)}_r=\bm W_a\bm F_r,
    \label{eq:dcrv2_extra_components}
\end{align}
where \(\bm W_d\in\mathbb{R}^{N_a\times d}\) and \(\bm W_a\in\mathbb{R}^{N_t\times d}\) are shared across the \(R_{\mathrm{e}}\) residual components. \textcolor{black}{The shared factorization provides a lightweight means of increasing the reconstruction order without assigning an independent high-dimensional projection to every additional component.} Each component is then weighted by a per-rank sigmoid gate \(g_r(\bm z)\in[0,1]\), inspired by the channel-wise attention mechanism in SENet~\cite{hu2018squeeze}, which adaptively re-weights feature channels via a sigmoid excitation. Here, the same idea is applied along the rank dimension, so that each residual component is treated as a rank channel whose contribution is emphasized or suppressed through end-to-end learning. Moreover, a global learnable scalar \(\alpha\) is introduced to control the overall strength of the supplementary components relative to the direct branch. The hybrid-rank CSI estimate is obtained by summing the direct reconstruction and the supplementary residual reconstruction, given by
\begin{align}
    \hat{\bm H}_a^{(\mathrm{hyb})}
    =
    \hat{\bm H}_a^{(\mathrm{dir})}
    +
    \alpha\sum_{r=1}^{R_{\mathrm{e}}} g_r(\bm z)\,\hat{\bm d}^{(e)}_r\left(\hat{\bm a}^{(e)}_r\right)^{H},
    \label{eq:dcrv2_hybrid_decoder}
\end{align}
where \(\alpha\) and the bias of \(g_r(\cdot)\) are initialized to a small value so that the residual branch starts near zero and the hybrid estimate falls back to \(\hat{\bm H}_a^{(\mathrm{dir})}\) at initialization. In this way, the direct components in \(\hat{\bm H}_a^{(\mathrm{dir})}\) carry the dominant multipath energy, while the gated bank in \eqref{eq:dcrv2_hybrid_decoder} acts as a switchable rank tail: the per-rank gates close on well-conditioned channels and open only on samples whose residual rank actually reduces the reconstruction error, while the global \(\alpha\) lets the network learn the overall trade-off between the two paths. In addition, the cost savings of this design are explicit. Since \(\bm W_d\) and \(\bm W_a\) are shared and \(\bm F_r\) is generated through the \(d\)-dimensional bottleneck, the factored form in \eqref{eq:dcrv2_extra_components} only costs \(\mathcal{O}(R_{\mathrm{e}}d+d(N_a+N_t))\) parameters, instead of the \(\mathcal{O}(R_{\mathrm{e}}M(N_a+N_t))\) required by enlarging the direct branch from \(R\) to \(R+R_{\mathrm{e}}\). Therefore, the effective reconstruction rank increases at a cost essentially independent of \(M\).

\begin{table*}[!ht]
  \centering
  \begin{threeparttable}
  \caption{Comparison of NMSE (dB) and FLOPs of model-based CSI feedback methods at compression ratio $\eta\in\{1/4,1/8,1/16,1/32\}$. The best result is in \textbf{bold} and the second best is \underline{underlined}; ``--'' denotes results not reported. Rows highlighted in color are the proposed methods.}
  \label{tab:1a}
  \cvprwidetablestyle
  \begin{tabularx}{\textwidth}{@{}l *{12}{Y}@{}}
  \toprule
  \multirow{2}{*}{\textbf{Method}}
      & \multicolumn{3}{c}{$\bm{\eta=1/4}$}
      & \multicolumn{3}{c}{$\bm{\eta=1/8}$}
      & \multicolumn{3}{c}{$\bm{\eta=1/16}$}
      & \multicolumn{3}{c}{$\bm{\eta=1/32}$} \\
  \cmidrule(lr){2-4}\cmidrule(lr){5-7}\cmidrule(lr){8-10}\cmidrule(lr){11-13}
      & FLOPs & Indoor & Outdoor
      & FLOPs & Indoor & Outdoor
      & FLOPs & Indoor & Outdoor
      & FLOPs & Indoor & Outdoor \\
  \midrule
  MS4L2O~\cite{ms4l2o}
    & --     & -5.96  & --
    & 21.50M & -5.61  & --
    & 10.75M & -4.11  & --
    & 5.38M  & -1.62  & -- \\
  ISTA~\cite{fista}
    & --     & -4.27  & --
    & 11.01M & -3.10  & --
    & 5.51M  & -1.47  & --
    & 2.75M  & -0.52  & -- \\
  FISTA~\cite{fista}
    & 42.99M & -10.46 & -6.35
    & 21.49M & -6.39  & -2.91
    & 10.75M & -3.18  & -1.15
    & 5.37M  & -1.11  & -0.35 \\
  \midrule
  \proposedrow LRP ($R_{\mathrm{enc}}=128$, $R=4$)
    & \textbf{0.68M} & -13.18 & -4.83
    & \textbf{0.48M} & -12.19 & -4.32
    & \textbf{0.39M} & -11.41 & -3.66
    & \textbf{0.34M} & -8.58  & -2.40 \\
  \proposedrow LRP ($R_{\mathrm{enc}}=512$, $R=4$)
    & \underline{1.88M} & -14.11 & -5.96
    & \underline{1.49M} & -14.15 & -5.30
    & \underline{1.29M} & -11.82 & \underline{-4.00}
    & \underline{1.20M} & -8.61  & -2.64 \\
  \proposedrow LRP ($R_{\mathrm{enc}}=512$, $R=8$)
    & 2.16M & \underline{-17.94} & \underline{-8.33}
    & 1.64M & \underline{-15.81} & \underline{-6.14}
    & 1.38M & \underline{-12.47} & \textbf{-4.19}
    & 1.25M & \underline{-8.62}  & \underline{-2.70} \\
  \proposedrow LRP ($R_{\mathrm{enc}}=512$, $R=16$)
    & 2.72M & \textbf{-21.62} & \textbf{-9.04}
    & 1.93M & \textbf{-15.87} & \textbf{-6.30}
    & 1.54M & \textbf{-12.55} & \textbf{-4.19}
    & 1.34M & \textbf{-8.67}  & \textbf{-2.72} \\
  \bottomrule
  \end{tabularx}
  \end{threeparttable}
  \end{table*}

\begin{table*}[!ht]
  \centering
  \begin{threeparttable}
  \caption{Comparison of NMSE (dB) and FLOPs of lightweight DL-based CSI feedback methods at compression ratio $\eta\in\{1/4,1/8,1/16,1/32\}$. The best result is in \textbf{bold} and the second best is \underline{underlined}. Rows highlighted in color are the proposed methods.}
  \label{tab:1b}
  \cvprwidetablestyle
  \begin{tabularx}{\textwidth}{@{}l *{12}{Y}@{}}
  \toprule
  \multirow{2}{*}{\textbf{Method}}
      & \multicolumn{3}{c}{$\bm{\eta=1/4}$}
      & \multicolumn{3}{c}{$\bm{\eta=1/8}$}
      & \multicolumn{3}{c}{$\bm{\eta=1/16}$}
      & \multicolumn{3}{c}{$\bm{\eta=1/32}$} \\
  \cmidrule(lr){2-4}\cmidrule(lr){5-7}\cmidrule(lr){8-10}\cmidrule(lr){11-13}
      & FLOPs & Indoor & Outdoor
      & FLOPs & Indoor & Outdoor
      & FLOPs & Indoor & Outdoor
      & FLOPs & Indoor & Outdoor \\
  \midrule
  CsiNet~\cite{CsiNet}
    & 5.41M & -17.36 & -8.75
    & 4.37M & -12.70 & -7.65
    & 3.84M & -8.65  & -4.51
    & 3.58M & -6.24  & -2.81 \\
  CRNet~\cite{CRNet}
    & 5.12M & -26.99 & -12.70
    & 4.07M & -16.01 & -8.04
    & 3.55M & -11.35 & -5.44
    & 3.28M & -8.93  & -3.51 \\
  CLNet~\cite{CLNet}
    & 4.05M & -29.16 & \underline{-12.88}
    & 3.01M & -15.60 & \underline{-8.29}
    & 2.48M & -11.15 & -5.56
    & 2.22M & -9.25  & \underline{-3.75} \\
  ACRNet-$1\times$~\cite{ACRNet}
    & 4.64M & -27.16 & -10.71
    & 3.60M & -15.34 & -7.85
    & 3.07M & -10.36 & -5.19
    & 2.81M & -8.60  & -3.31 \\
  DCRNet-$1\times$~\cite{DCRNet}
    & 4.01M & -28.04 & -12.58
    & 2.96M & -16.26 & -7.95
    & 2.44M & -11.74 & \underline{-5.60}
    & 2.18M & -9.05  & -3.47 \\
  \midrule

  \proposedrow DCRNetV2-mini
    & \textbf{2.83M} & -31.00         & -10.83
    & \textbf{1.87M} & -19.23         & -7.20
    & \textbf{1.39M} & -14.18         & -4.83
    & \textbf{1.15M} & \underline{-9.55}  & -3.14 \\

    \proposedrow DCRNetV2-small
    &\underline{4.00M} & -31.12         & -11.44
    & \underline{2.67M} & \textbf{-20.51} & -7.70
    & \underline{2.05M} & \underline{-14.64} & -5.17
    & \underline{1.74M} & -9.16          & -3.31 \\

    \proposedrow DCRNetV2-base
    & 5.23M & \underline{-32.75} & -11.89
    & 3.88M & -20.07         & -7.83
    & 3.20M & \textbf{-14.73} & -5.51
    & 2.86M & -9.35          & -3.56 \\

    \proposedrow DCRNetV2-unified
    & 6.59M & \textbf{-32.84} & \textbf{-12.99}
    & 4.64M & \underline{-20.36} & \textbf{-8.65}
    & 3.66M & -14.62         & \textbf{-5.85}
    & 3.17M & \textbf{-9.69}  & \textbf{-3.79} \\

  %   \proposedrow DCRNet-LRP-1$\times$  ($r=32$, $d=512$)
  %   & 5.35M & \underline{-30.97} & -12.16
  %   & 3.51M & -17.87 & \textbf{-8.43}
  %   & 2.60M & \underline{-13.80}     & -5.50
  %   & 2.17M & \underline{-9.21}       &  -3.22\\
  % \proposedrow DCRNet-LRP-1$\times$ ($r=16$, $d=1024$)
  %   & 5.84M & \textbf{-31.70} & -11.41
  %   & 4.27M & \textbf{-19.27} & -7.94
  %   & 3.48M & -13.78          & -5.15
  %   & 3.09M & -9.15              & -3.40 \\
  \bottomrule
  \end{tabularx}
  \end{threeparttable}
  \end{table*}

\subsection{DCR Refinement Module}
\label{subsec:dcrnetv2_refine}
\textcolor{black}{After the hybrid-rank decoder restores the codeword to the full angular-delay domain, we further apply a DCR refinement module sequentially to the reconstructed CSI. Unlike the encoder, where local features must be extracted before dimensionality reduction, the decoder has already recovered a full-dimensional channel estimate at this stage. Therefore, the refinement module can directly exploit the spatial neighborhood of each angular-delay coefficient without passing through another information bottleneck. This naturally leads to a reconstruction-then-refinement structure, which is also commonly adopted in CSI feedback networks~\cite{CsiNet,CsiNetPlus,DCRNet}.}

As shown in Fig.~\ref{fig:dcrnetv2_arch}, this module contains two parallel convolutional paths followed by concatenation and a \(1\times1\) fusion layer. One path uses dilated and asymmetric kernels to collect local off-grid leakage over a larger angular-delay neighborhood, while the other path preserves short-range details with standard and asymmetric convolutions. Denoting this single refinement operator by \(f_{\mathrm{ref}}(\cdot)\), the refined estimate is
\begin{align}
    \hat{\bm H}_a^{(\mathrm{ref})}
    =
    \hat{\bm H}_a^{(\mathrm{hyb})}
    +
    \gamma_{\mathrm{ref}}\,
    f_{\mathrm{ref}}\!\left(\hat{\bm H}_a^{(\mathrm{hyb})}\right),
    \label{eq:dcrv2_refine_block}
\end{align}
where \(\gamma_{\mathrm{ref}}\) is a learnable scalar gate initialized close to zero. 

\subsection{Complexity Control}
\label{subsec:dcrnetv2_variants}
To balance reconstruction accuracy and computational cost, we introduce several DCRNetV2 variants. These models share the same architecture but vary across four parameters: the sensing rank $R_{\mathrm{enc}}$ at the UE, and the direct rank $R$, residual rank $R_{\mathrm{e}}$, and bottleneck size $d$ at the BS.At the UE, $R_{\mathrm{enc}}$ denotes the number of rank-one sensing operators with complexity $\mathcal{O}(R_{\mathrm{enc}}(N_aN_t + M))$. Reducing $R_{\mathrm{enc}}$ directly lowers the encoder FLOPs for user devices. At the BS, the direct rank $R$ sets the number of main components in \eqref{eq:dcrv2_direct_decoder}, where a small $R$ captures dominant path energy and a larger $R$ provides a finer low-rank base. The residual branch then introduces $R_{\mathrm{e}}$ extra components at a modest cost of $\mathcal{O}(R_{\mathrm{e}}d(N_a + N_t))$, expanding the effective rank without running full high-dimensional projections.Adjusting these four parameters yields a spectrum of models, from compact user-friendly setups to high-accuracy variants, all built upon the same low-rank backbone. Section~\ref{sec:simulation} details the specific configurations used in our experiments.

\begin{table}[t]
  \centering
  \begin{threeparttable}
  \caption{Comparison of high-performance DL-based CSI feedback methods at $\eta\in\{1/4,1/16\}$. The best result is in \textbf{bold} and the second best is \underline{underlined}; ``--'' denotes results not reported. Rows highlighted in color are the proposed methods.}
  \label{tab:high_perf}
  \cvprnarrowtablestyle
  \begin{tabularx}{\columnwidth}{@{}l *{6}{Y}@{}}
  \toprule
  \multirow{2}{*}{\textbf{Method}}
      & \multicolumn{3}{c}{$\bm{\eta=1/4}$}
      & \multicolumn{3}{c}{$\bm{\eta=1/16}$} \\
  \cmidrule(lr){2-4}\cmidrule(lr){5-7}
      & FLOPs & Indoor & Outdoor
      & FLOPs & Indoor & Outdoor \\
  \midrule
  DS-NLCsiNet~\cite{DSNLCsiNet}
    & 11.30M & -24.99 & -12.09
    & 9.72M  & -12.93 & -4.98 \\
  CsiNet$+$~\cite{CsiNetPlus}
    & 24.57M & -27.37 & -12.40
    & 23.00M & -14.14 & -5.73 \\
  ACRNet-$10\times$~\cite{ACRNet}
    & 24.40M & -29.83 & -13.61
    & 22.82M & -14.32 & -6.30 \\
  DCRNet-$10\times$~\cite{DCRNet}
    & 17.57M & -30.61 & -13.72
    & 16.00M & -14.02 & -6.35 \\
  TransNet~\cite{TransNet}
    & 35.72M & -32.38 & \textbf{-14.86}
    & 34.14M & \textbf{-15.00} & \textbf{-7.82} \\
  TCLNet~\cite{TCLNet}
    & 31.47M & -29.29 & --
    & 29.89M & -14.77 & -- \\
  LLMCsiNet-S~\cite{LLMCsiNet}
    & 19.85M & --     & --
    & 19.85M & -14.25 & -- \\
  SSE-Net~\cite{SSE-Net}
    & 10.37M & -32.06 & -13.73
    & 8.88M  & -13.39 & -6.03 \\
  \midrule
  \proposedrow DCRNetV2-large-O
    & \underline{8.25M} & -- & \underline{-14.85}
    & \textbf{3.93M} & -- & \underline{-6.81} \\
  \proposedrow DCRNetV2-large-I
    & \textbf{7.00M} & \textbf{-33.20} & --
    & \underline{3.97M} & \underline{-14.79} & -- \\
  \bottomrule
  \end{tabularx}
  \end{threeparttable}
\end{table}

\section{Simulation Results}\label{sec:simulation}
In this section, we present the simulation results of the proposed LRP and DCRNetV2 methods, and compare them with representative model-based and DL-based CSI feedback methods in terms of reconstruction accuracy and computational cost. We also analyze the trade-off between accuracy and complexity, and evaluate the inference latency and energy consumption on a real mobile device. Moreover, we test the generalization performance of the proposed methods under the proposed benchmark and real measured CSI datasets.

\subsection{General Setup}
\label{subsec:sim_setup}
Following the standard practice in \cite{CsiNet,CsiNetPlus}, we first evaluate the proposed methods on the COST 2100 channel dataset~\cite{cost2100} under the indoor and outdoor scenarios, with 5.3 GHz and 300 MHz carrier frequencies, respectively. The BS is equipped with \(N_t=32\) antennas and \(N_c=1024\) OFDM subcarriers, and the truncated angular-delay CSI has \(N_a=32\) rows. The training, validation, and test sets contain \(100{,}000\), \(30{,}000\), and \(20{,}000\) samples, respectively.  We report the normalized mean squared error (NMSE) in dB, given by
\begin{align}
    \mathrm{NMSE} = 10\log_{10}\left(\mathbb E\left[\frac{\|\hat{\bm H}_a-\bm H_a\|_F^2}{\|\bm H_a\|_F^2}\right]\right).
\end{align}
We also evaluate the computational cost in terms of FLOPs to compare the efficiency of different methods. For the proposed LRP and DCRNetV2 variants, we train each model for 1500 epochs using mean squared error loss and the Adam optimizer with a batch size of 200 and an initial learning rate of \(2\times10^{-3}\).

\begin{figure}[t!]
  \centering
  \includegraphics[width=\columnwidth]{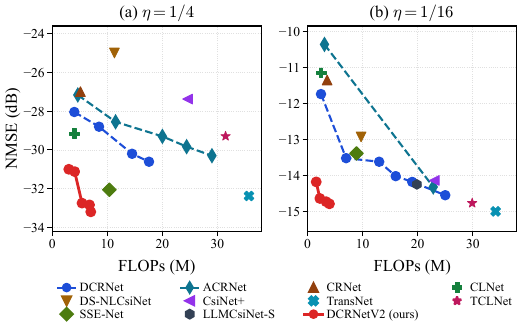}
  \caption{FLOPs vs.\ indoor NMSE trade-off for representative DL-based CSI feedback methods at (a) $\eta=1/4$ and (b) $\eta=1/16$.}
  \label{fig:flops_vs_nmse}
\end{figure}

\begin{figure}[t]
  \centering
  \includegraphics[width=\columnwidth]{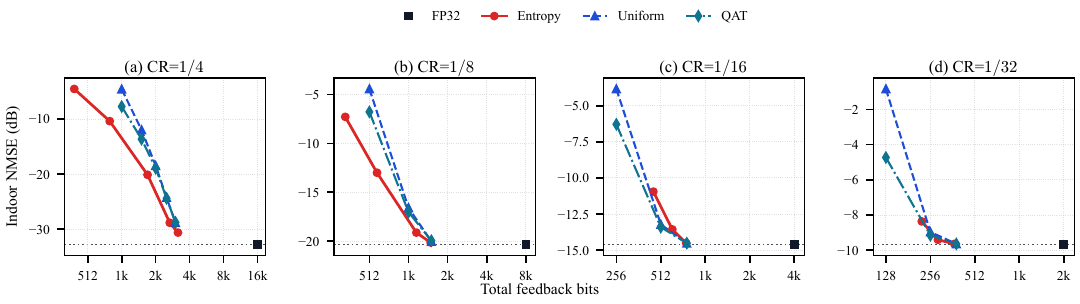}
  \caption{Indoor quantization performance versus total feedback bits under different compression ratios. The FP32 point and dotted line indicate the unquantized reference for each compression ratio.}
  \label{fig:indoor_quant_bits_nmse}
\end{figure}
\subsection{Baseline Methods}
\label{subsec:dcrnetv2_variants_sim}
To evaluate the effectiveness of the proposed LRP architecture, we compare it with representative model-based feedback methods, including MS4L2O~\cite{ms4l2o}, the classical iterative shrinkage thresholding algorithm (ISTA), and its accelerated variant FISTA~\cite{fista}. To evaluate the proposed DCRNetV2 architecture, we first compare it with lightweight DL-based feedback networks, including CsiNet~\cite{CsiNet}, CRNet~\cite{CRNet}, CLNet~\cite{CLNet}, ACRNet-$1\times$ and DCRNet-$1\times$~\cite{DCRNet}. We also compare with high-performance DL-based feedback methods, including DS-NLCsiNet~\cite{DSNLCsiNet}, CsiNet$+$~\cite{CsiNetPlus}, ACRNet-$10\times$ and DCRNet-$10\times$~\cite{DCRNet}, TransNet~\cite{TransNet}, TCLNet~\cite{TCLNet}, LLMCsiNet-S~\cite{LLMCsiNet}, and SSE-Net~\cite{SSE-Net}.

We implement standalone LRP by varying the encoder sensing rank \(R_{\mathrm{enc}}\in\{128,512\}\) and the decoder synthesis rank \(R\in\{4,8,16\}\) to examine the low-rank sensing and reconstruction trade-off. For DCRNetV2, we use five variants with gradually increased capacity. The mini version adopts \((R_{\mathrm{enc}},R,R_{\mathrm{e}},d)=(128,10,8,8)\), where \(R_{\mathrm{enc}}\) denotes the encoder rank, \(R\) is the number of direct decoder components, \(R_{\mathrm{e}}\) is the number of factored residual components, and \(d\) is only the residual-decoder bottleneck dimension. The small and base versions remove the residual branch and use a plain low-rank decoder with \(R=16\), while increasing \(R_{\mathrm{enc}}\) from \(256\) to \(512\). The unified version uses \((R_{\mathrm{enc}},R,R_{\mathrm{e}},d)=(512,32,16,16)\), and the large version further increases the residual rank to \(R_{\mathrm{e}}=32\) while keeping \(R_{\mathrm{enc}}=512\), \(R=32\), and \(d=16\).

\subsection{Performance Comparison}
\label{subsec:sim_baselines}
In Table~\ref{tab:1a}, we first compare the standalone LRP architecture with model-based feedback methods under the COST 2100 indoor and outdoor scenarios. From this table, we can see that LRP consistently outperforms the model-based methods in terms of NMSE for the same or lower FLOPs. For example, at \(\eta=1/4\), LRP with \(R_{\mathrm{enc}}=128\) and \(R=4\) achieves \(-13.18\)~dB indoor NMSE with only \(0.68\)M FLOPs, while the best competitor FISTA gives \(-10.46\)~dB with \(42.99\)M FLOPs. The trend is similar for another compression ratio and the outdoor scenario. It is worth noting that when the encoder sensing rank \(R_{\mathrm{enc}}\) and decoder synthesis rank \(R\) are increased, LRP can further improve the NMSE while still maintaining a much lower FLOPs than the model-based methods, even outperforming or approaching the performance of the lightweight DL-based feedback methods in Table~\ref{tab:1b}. This is because the proposed LRP architecture keeps the physical low-rank prior in the compression and reconstruction process, which is more effective than the generic random projection and iterative recovery used in the model-based methods. These results validate the effectiveness of the proposed LRP architecture.

\begin{figure}[t]
  \centering
  \includegraphics[width=0.85\columnwidth]{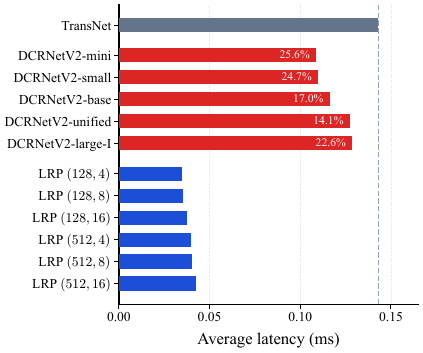}
  \caption{Average inference latency on iPhone 17 Pro Max using Core ML. The dashed line marks TransNet. Percentages on the DCRNetV2 bars are the energy reduction relative to TransNet. For LRP, $(R_{\mathrm{enc}}, R)$ is given in parentheses.}
  \label{fig:iphone_latency}
\end{figure}

Table~\ref{tab:1b} compares DCRNetV2 with lightweight DL baselines under $\sim$5M FLOPs at $\eta=1/4$. Across this low-complexity regime, DCRNetV2 offers a superior accuracy-complexity trade-off. Specifically, at $\eta=1/4$, DCRNetV2-mini lowers the indoor NMSE from $-28.04$~dB (DCRNet-$1\times$) to $-31.00$~dB while cutting FLOPs from $4.01$M to $2.83$M. At $\eta=1/8$, it reaches nearly $-20$~dB NMSE with only $1.87$M FLOPs, roughly half the computation of ACRNet-$1\times$ with a $>4$~dB gain. Scaled variants further enhance accuracy; for instance, DCRNetV2-unified achieves $-32.84$~dB indoor NMSE at $\eta=1/4$ and SOTA outdoor performance with modest FLOP increases.

Table~\ref{tab:high_perf} further evaluates DCRNetV2 against high-performance, large Transformer-based baselines. Scenario-specific variants DCRNetV2-large-I and DCRNetV2-large-O achieve competitive accuracy with substantially lower complexity. At $\eta=1/4$, DCRNetV2-large-I reaches $-33.20$~dB indoor NMSE with $7.00$M FLOPs, outperforming TransNet's $-32.38$~dB at only one-fifth of the computation. For outdoor channels, DCRNetV2-large-O matches TransNet with $-14.85$~dB at $8.25$M FLOPs. Even at $\eta=1/16$, where TransNet holds a marginal NMSE lead, both large variants yield comparable accuracy while slashing FLOPs by nearly $10\times$. These results demonstrate that DCRNetV2 effectively scales into the high-accuracy regime with superior computational efficiency.

Fig.~\ref{fig:indoor_quant_bits_nmse} evaluates the actual feedback-bit cost after quantizing the DCRNetV2 indoor codeword. The FP32 markers provide the unquantized reference, while entropy coding, uniform quantization, and quantization-aware training (QAT) are compared under the same compression ratios. Entropy coding preserves most of the FP32 accuracy at moderate bit widths with far fewer transmitted bits. For example, at \(\eta=1/8\), \(\eta=1/16\), and \(\eta=1/32\), the 6-bit entropy-coded codewords require only \(1513\), \(760\), and \(372\) bits, respectively, while staying close to the FP32 references. At \(\eta=1/4\), entropy coding reduces the feedback load from \(16384\) bits to \(3228\) bits, with NMSE changing from \(-32.84\)~dB to \(-30.65\)~dB. Under very low bit widths, entropy coding degrades more gracefully than uniform quantization and QAT, indicating that the learned codeword has a non-uniform symbol distribution that should be exploited by the feedback quantizer.

Fig.~\ref{fig:iphone_latency} further shows the on-device inference latency and energy reduction on iPhone 17 Pro Max using the Core ML framework\footnote{\url{https://developer.apple.com/machine-learning/core-ml/}}. Since Transformer models have recently become dominant in AI community, Core ML provides mature graph and kernel optimizations for them, from which TransNet can directly benefit. In contrast, the proposed LRP and DCRNetV2 rely on less standard asymmetric separable convolutions that are not equally optimized by the current backend. Even under this setting, standalone LRP requires only \(0.0348\)--\(0.0423\)~ms on average, which is about \(3.4\)--\(4.1\) times faster than TransNet. After adding the DCRNetV2 residual modules, all listed variants still have lower average latency than TransNet. For example, DCRNetV2-mini achieves \(0.1088\)~ms average latency with \(25.6\%\) energy reduction. These results show that the proposed approaches remain efficient on real mobile hardware, even without backend optimizations.

\subsection{Ablation Studies}
\label{subsec:sim_ablation}
\begin{table}[t]
\centering
\caption{Ablation study of different architectural components on the COST 2100 indoor scenario at \(\eta=1/4\).}
\label{tab:ablation}
\footnotesize
\renewcommand{\arraystretch}{1.08}
\setlength{\tabcolsep}{2pt}
\begin{tabularx}{\columnwidth}{@{}YYYYYY@{}}
\toprule
\multicolumn{2}{c}{\textbf{Enc.}}
& \multicolumn{2}{c}{\textbf{Dec.}}
& \multirow{2}{*}{\textbf{NMSE}}
& \multirow{2}{*}{\textbf{FLOPs}} \\
\cmidrule(lr){1-2} \cmidrule(lr){3-4}
\textbf{LRP} & \textbf{DCR}
& \textbf{LRP} & \textbf{DCR}
&  &  \\
\midrule
\cmark & \xmark & \cmark & \xmark & -21.62 & 2.72M \\
\cmark & \xmark & \cmark & \cmark & -29.34 & 3.84M \\
\cmark & \cmark & \cmark & \xmark & \underline{-32.00} &  6.45M \\
\proposedrow \cmark & \cmark & \cmark & \cmark & \textbf{-32.84} & 6.59M \\
\bottomrule
\end{tabularx}
\end{table}

\begin{table}[t]
\centering
\caption{Effect of the low-rank prior across different backbones on the indoor scenario.
NMSE is reported in dB, where lower is better. \(\Delta\) denotes the NMSE improvement brought by the low-rank prior.}
\label{tab:lr_prior_ablation}
\footnotesize
\renewcommand{\arraystretch}{1.08}
\setlength{\tabcolsep}{2.5pt}
\begin{threeparttable}
\begin{tabularx}{\columnwidth}{@{}lcYYYY@{}}
\toprule
\multirow{2}{*}{\textbf{Backbone}}
& \multirow{2}{*}{\textbf{LRP}}
& \multicolumn{2}{c}{\textbf{\(\bm{\eta=1/4}\)}}
& \multicolumn{2}{c}{\textbf{\(\bm{\eta=1/16}\)}} \\
\cmidrule(lr){3-4} \cmidrule(lr){5-6}
& & \textbf{NMSE} & \(\boldsymbol{\Delta}\)
  & \textbf{NMSE} & \(\boldsymbol{\Delta}\) \\
\midrule
\multirow{2}{*}{CsiNet\tnote{a}}
& \xmark & -22.24 & --    & -11.25 & -- \\
& \cmark & \cellcolor{proposedblue}\textbf{-27.90} & \cellcolor{proposedblue}-5.66 & \cellcolor{proposedblue}\textbf{-13.02} & \cellcolor{proposedblue}-1.78 \\
\midrule
\multirow{2}{*}{Conv5-MLP}
& \xmark & -28.05 & --    &  -8.99 & -- \\
& \cmark & \cellcolor{proposedblue}\textbf{-29.08} & \cellcolor{proposedblue}-1.03 & \cellcolor{proposedblue}\textbf{-13.20} & \cellcolor{proposedblue}-4.22 \\
\bottomrule
\end{tabularx}
\begin{tablenotes}
\footnotesize
\item[a] We reimplemented CsiNet in PyTorch because the original TensorFlow code is no longer maintained, which may lead to differences from the results reported in~\cite{CsiNet}.
\end{tablenotes}
\end{threeparttable}
\end{table}

\begin{figure*}[t!]
  \centering
  \includegraphics[width=0.9\textwidth]{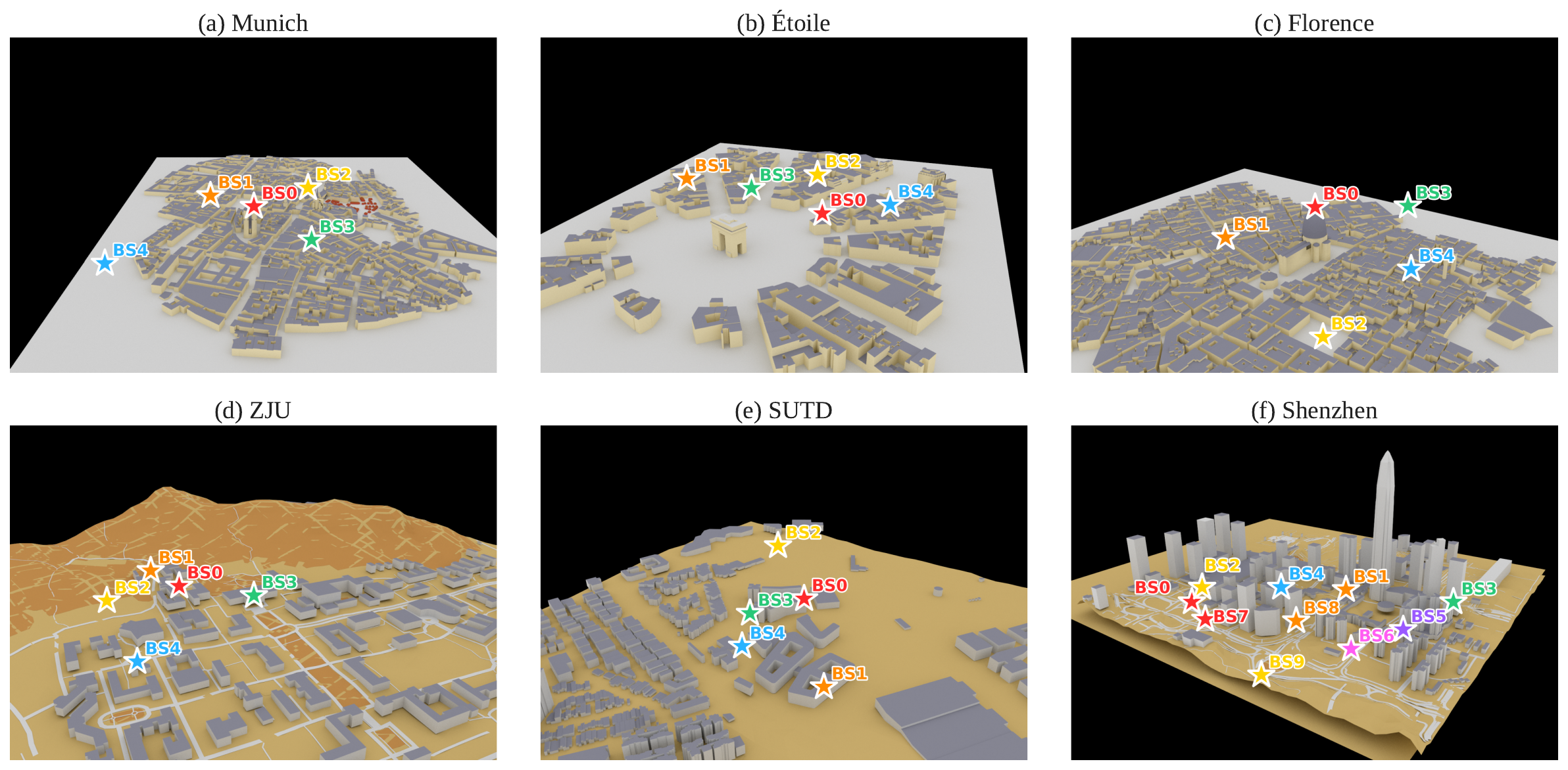}
  \caption{Outdoor scenes and BS placements in the Sionna-RT-Mix5 dataset. The scenes cover European urban layouts and campus-like deployments with different LoS and NLoS characteristics.}
  \label{fig:sionna_scenarios}
\end{figure*}

\begin{table*}[t]
  \centering
  \begin{threeparttable}
  \caption{NMSE (dB) comparison of different CSI feedback methods on Sionna-RT-Mix5 dataset, where compression ratio is set to \(\eta\in\{1/4,1/8,1/16\}\). The best result in each column is in \textbf{bold} and the second-best is \underline{underlined}. Rows highlighted in color are the proposed methods.}
  \label{tab:rt_csi_scene_cr}
  \cvprwidetablestyle
  \begin{tabularx}{\textwidth}{@{}l *{15}{Y}@{}}
  \toprule
  \multirow{2}{*}{\textbf{Method}}
       & \multicolumn{3}{c}{\textbf{Paris}}
        & \multicolumn{3}{c}{\textbf{Florence}}
      & \multicolumn{3}{c}{\textbf{Munich}}
      & \multicolumn{3}{c}{\textbf{ZJU}}
      & \multicolumn{3}{c}{\textbf{SUTD}} \\
  \cmidrule(lr){2-4}\cmidrule(lr){5-7}\cmidrule(lr){8-10}\cmidrule(lr){11-13}\cmidrule(lr){14-16}
      & \(\bm{1/4}\) & \(\bm{1/8}\) & \(\bm{1/16}\)
      & \(\bm{1/4}\) & \(\bm{1/8}\) & \(\bm{1/16}\)
      & \(\bm{1/4}\) & \(\bm{1/8}\) & \(\bm{1/16}\)
      & \(\bm{1/4}\) & \(\bm{1/8}\) & \(\bm{1/16}\)
      & \(\bm{1/4}\) & \(\bm{1/8}\) & \(\bm{1/16}\) \\
  \midrule

    \proposedrow  LRP
      & -17.01 & -14.21 & -12.20
      & -19.31 & -15.81 & -13.27
      & -16.18 & -13.12 & -10.85
      & -14.86 & -12.48 & -10.43
      & -14.58 & -12.08 & -10.69 \\   \midrule
  CRNet
      & -18.65 & -15.56 & -13.75
      & -20.77 & -17.39 & -15.37
      & -18.41 & -14.93 & -12.79
      & -16.29 & -13.41 & -12.11
      & -15.98 & -13.39 & -12.01 \\
    CLNet
      & \underline{-19.92} & -16.88 & \underline{-14.26}
      & \underline{-23.14} & \underline{-19.52} & \underline{-16.23}
      & \underline{-19.84} & \textbf{-16.34} & \underline{-13.24}
      & \underline{-16.77} & -14.44 & -12.20
      & -16.36 & -14.40 & -12.17 \\
    DCRNet-1\(\times\)
      & -13.49 & -13.76 & -12.61
      & -14.59 & -15.01 & -13.94
      & -12.34 & -12.81 & -11.51
      & -11.21 & -12.04 & -10.85
      & -11.22 & -11.98 & -10.77 \\
    TransNet
      & -19.32 & \textbf{-17.52} & \textbf{-15.86}
      & -22.41 & \textbf{-19.82} & \textbf{-17.47}
      &	-18.74  & \underline{-16.25} & \textbf{-14.25}
      & -16.58 & \textbf{-15.70} & \textbf{-13.98}
      & \underline{-16.93} & \textbf{-15.45} & \textbf{-14.00} \\
  \midrule
  \proposedrow DCRNetV2-unified
      & \textbf{-20.63} & \underline{-16.89} & -14.10
      & \textbf{-24.40} & -19.50 & -15.86
      & \textbf{-20.16} & -15.90 & -12.67
      & \textbf{-17.82} & \underline{-14.62} & \underline{-12.24}
      & \textbf{-17.53} & \underline{-14.61} & \underline{-12.28} \\
  \bottomrule
  \end{tabularx}
  \end{threeparttable}
  \end{table*}

In Table~\ref{tab:ablation}, we take DCRNetV2-unified as the full model and conduct ablation studies to analyze the contribution of different components. The first row corresponds to the pure LRP backbone without any residual paths, which already achieves a reasonable NMSE of \(-21.62\)~dB with only \(2.72\)M FLOPs. Adding the decoder-side DCR branch improves the NMSE to \(-29.34\)~dB with \(3.84\)M FLOPs, showing the benefit of refinement after low-rank synthesis. Adding the DCR encoder without the decoder residual reaches \(-32.00\)~dB, which indicates that convolutional user-side features also help preserve off-grid details in the feedback codeword. These results validate the design of the DCRNetV2 architecture, where the LRP backbone captures the dominant low-rank component while the residual branches compensate for the remaining details that lie outside the low-rank subspace.

To further verify the generality of the low-rank prior, we conduct an ablation study by adding the LRP path to two different convolutional backbones: the CsiNet architecture and a simple convolutional encoder with size \(5\times5\) kernels followed by an MLP decoder. From this table, we can see that adding the LRP path consistently improves the NMSE for both backbones at different compression ratios. For CsiNet, LRP brings a significant improvement of \(5.66\)~dB at \(\eta=1/4\) and \(1.78\)~dB at \(\eta=1/16\). For the Conv5-MLP backbone, the gains are \(1.03\)~dB and \(4.22\)~dB, respectively. These results support the effectiveness of the low-rank prior in improving CSI feedback performance, and show that the proposed LRP architecture has potential to enhance various backbone designs.

\subsection{Extension to More Scenarios}
\label{subsec:sim_generalization}

To further validate the generalization performance of the proposed methods, we introduce a new ray-tracing CSI feedback benchmark. Specifically, we use NVIDIA Sionna\footnote{\url{https://developer.nvidia.com/sionna}} to generate CSI samples from five outdoor scenes (Sionna-RT-Mix5) in Fig.~\ref{fig:sionna_scenarios}. Among them, Paris, Florence, and Munich are public scenes released by NVIDIA, while Zhejiang University (ZJU) and Singapore University of Technology and Design (SUTD) are self-constructed campus-scale scenes to cover different LoS and NLoS propagation characteristics. We also conduct the Shenzhen CBD skyline scenario for test-only evaluation. More details can be found in Appendix~\ref{app:rt_scenarios}\footnote{The dataset can be accessed at \url{https://github.com/tangshunpu/RT-CSI}.}.

Table~\ref{tab:rt_csi_scene_cr} reports the corresponding NMSE results at \(\eta\in\{1/4,1/8,1/16\}\) on the five Sionna-RT-Mix5 scenes. All models are trained on the training set of this dataset including all five scenes, and evaluated on the test set of each scene separately. From this table, we can see that DCRNetV2-unified achieves the best NMSE when $\eta=1/4$ in all five scenes, and is competitive with TransNet in the remaining cases. These results show that the proposed DCRNetV2 architecture can generalize well to more realistic ray-tracing channels with different propagation characteristics, and can achieve competitive performance compared with the SOTA DL-based feedback methods with much lower computational cost. 

\begin{figure}
  \centering
  \includegraphics[width=0.95\columnwidth]{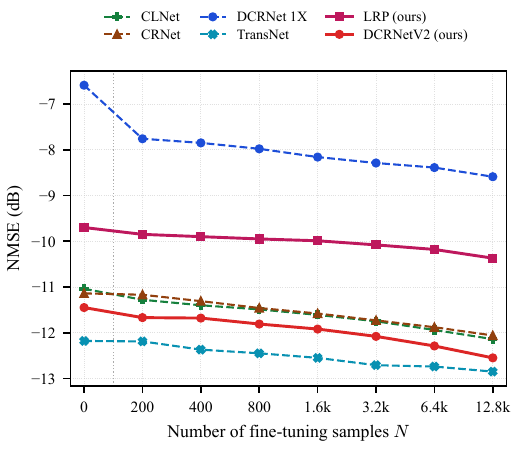}
  \caption{NMSE (dB) versus the number of fine-tuning samples for different CSI feedback methods on the Shenzhen Futian CBD scenario. All models are pretrained on Sionna-RT-Mix5, and \(N=0\) denotes evaluation without fine-tuning.}
  \label{fig:finetune_samples_nmse}
\end{figure}

We also evaluate out-of-distribution generalization on the Shenzhen CBD scenario. Specifically, all models are pretrained on Sionna-RT-Mix5 and tested on \(40{,}000\) Shenzhen samples without fine-tuning. We then fine-tune these models with a small number of Shenzhen training samples. From the results in Fig.~\ref{fig:finetune_samples_nmse}, we can see that DCRNetV2-unified achieves the second-best NMSE without fine-tuning, which is less than \(1\)~dB worse than TransNet. After fine-tuning, DCRNetV2-unified quickly approaches TransNet. These results indicate that, although the limited parameter count of DCRNetV2 may lead to slightly worse out-of-distribution NMSE than large Transformers, it still achieves competitive accuracy at much lower FLOPs and adapts to a new scenario with only a few fine-tuning samples.

In Table~\ref{tab:real_fdd}, we further compare the proposed methods with representative DL-based feedback methods on the real FDD Massive MIMO dataset \cite{directional_training,massive_mimo_channels} collected by the Renew group. From the table, we can see that the proposed DCRNetV2-unified achieves the best NMSE at all tested compression ratios. At \(\eta=1/16\), it obtains \(-19.04\)~dB, outperforming the second-best CLNet by \(2.11\)~dB. At \(\eta=1/32\), it improves over DCRNet-\(1\times\) by \(2.87\)~dB with comparable FLOPs. We note that the proposed LRP also achieves competitive NMSE, even outperforming some DL-based feedback methods at \(\eta=1/32\) and achieving the second best NMSE at \(\eta=1/64\). These results further confirm the effectiveness of the proposed methods on real measured channels.

\section{Conclusion}\label{sec:conclusion}
In this paper, we have investigated CSI feedback by exploiting the low-rank multipath structure of the channel. Specifically, we proposed the LRP architecture, which learns structured sensing and reconstruction over rank-one channel components. We further developed DCRNetV2 by combining the LRP backbone with convolutional residual branches to capture both the dominant low-rank components and the remaining channel details. Moreover, we introduced a ray-tracing CSI feedback benchmark with realistic outdoor scenarios to evaluate the generalization performance of different methods. Extensive experiments demonstrate that the proposed methods achieve a better accuracy--complexity trade-off than existing model-based and DL-based feedback methods and generalize well to ray-tracing and real-world scenarios. However, the path-wise rank-one formulation of LRP relies on a far-field channel model with approximately frequency-invariant array responses. In near-field wideband scenarios, spherical wavefronts and frequency-dependent spatial responses introduce angle--distance--frequency coupling, which may weaken the spatial--frequency separability of individual paths and limit the accuracy of the corresponding rank-one approximations~\cite{Cui2023NearFieldWideband}. Future work includes extending LRP to near-field wideband channels through frequency-dependent spherical-wave modeling and local low-rank representations over subbands and subarrays~\cite{Cui2024NearFieldBeamforming}. We will also explore additional physical priors and incorporate Transformer architectures to further improve CSI feedback performance.

\begin{table}[t]
  \centering
  \begin{threeparttable}
  \caption{Comparison of representative DL-based CSI feedback methods at \(\eta\in\{1/16,1/32,1/64\}\) on the real FDD Massive MIMO dataset of the Renew group.  The best result in each column is in \textbf{bold} and the second-best is \underline{underlined}. Rows highlighted in color are the proposed methods.}
  \label{tab:real_fdd}
  \cvprnarrowtablestyle
  \begin{tabularx}{\columnwidth}{@{}l *{6}{Y}@{}}
  \toprule
  \multirow{2}{*}{\textbf{Method}}
      & \multicolumn{2}{c}{\(\bm{\eta=1/16}\)}
      & \multicolumn{2}{c}{\(\bm{\eta=1/32}\)}
      & \multicolumn{2}{c}{\(\bm{\eta=1/64}\)} \\
  \cmidrule(lr){2-3}\cmidrule(lr){4-5}\cmidrule(lr){6-7}
      & \textbf{FLOPs} & \textbf{NMSE}
      & \textbf{FLOPs} & \textbf{NMSE}
      & \textbf{FLOPs} & \textbf{NMSE} \\
  \midrule
  \proposedrow LRP
      & \textbf{9.12M} & -15.51
      & \textbf{7.33M} & -15.68
      & \textbf{6.43M} & \underline{-14.59} \\
  \midrule
  CRNet
      & 28.92M          & -16.01
      & 22.69M          & -11.80
      & 19.58M          & -10.44 \\
  CLNet
      & 24.36M          & \underline{-16.93}
      & 18.13M          & -13.39
      & 15.02M          & -9.16 \\
  DCRNet-\(1\times\)
      & \underline{23.70M} & -16.72
      & \underline{17.47M} & \underline{-16.00}
      & \underline{14.36M} & -12.80 \\
  \midrule
  \proposedrow DCRNetV2-unified
      & 26.79M          & \textbf{-19.04}
      & 20.26M          & \textbf{-18.87}
      & 16.99M          & \textbf{-17.62} \\
  \bottomrule
  \end{tabularx}
  \end{threeparttable}
\end{table}

\bibliographystyle{IEEEtran}
\bibliography{IEEEabrv,references}

@STRING{IEEE_J_SP         = "{IEEE} Trans. Signal Processing"}

@STRING{IEEE_J_WCOM       = "{IEEE} Trans. Wirel. Commun."}

@STRING{IEEE_J_PAMI       = "{IEEE} Trans. Pattern Anal. Machine Intell."}

@article{Cui2023NearFieldWideband,
  author  = {Cui, Mingyao and Dai, Linglong},
  title   = {Near-field wideband channel estimation for extremely large-scale {MIMO}},
  journal = {Sci. China Inf. Sci.},
  volume  = {66},
  number  = {7},
  pages   = {172303},
  year    = {2023},
  doi     = {10.1007/s11432-022-3654-y}
}

@article{larsson2014massive,
  author  = {Erik G. Larsson and
             Ove Edfors and
             Fredrik Tufvesson and
             Thomas L. Marzetta},
  title   = {Massive {MIMO} for Next Generation Wireless Systems},
  journal = {{IEEE} Commun. Mag.},
  volume  = {52},
  number  = {2},
  pages   = {186--195},
  year    = {2014},
  doi     = {10.1109/MCOM.2014.6736761}
}

@article{sim2016compressed,
  author  = {Sim, Min Soo and
             Park, Jeonghun and
             Chae, Chan{-}Byoung and
             Heath, Robert W.},
  title   = {Compressed Channel Feedback for Correlated Massive {MIMO} Systems},
  journal = {J. Commun. Netw.},
  volume  = {18},
  number  = {1},
  pages   = {95--104},
  month   = feb,
  year    = {2016},
  doi     = {10.1109/JCN.2016.000012}
}

@article{rao2014distributed,
  author  = {Xiongbin Rao and
             Vincent K. N. Lau},
  title   = {Distributed Compressive {CSIT} Estimation and Feedback for {FDD} Multi-User Massive {MIMO} Systems},
  journal = {{IEEE} Trans. Signal Process.},
  volume  = {62},
  number  = {12},
  pages   = {3261--3271},
  month   = jun,
  year    = {2014},
  doi     = {10.1109/TSP.2014.2324991}
}

@article{love2008overview,
  author  = {David J. Love and
             Robert W. Heath and
             Vincent K. N. Lau and
             David Gesbert and
             Bhaskar D. Rao and
             Matthew Andrews},
  title   = {An Overview of Limited Feedback in Wireless Communication Systems},
  journal = {{IEEE} J. Sel. Areas Commun.},
  volume  = {26},
  number  = {8},
  pages   = {1341--1365},
  month   = oct,
  year    = {2008},
  doi     = {10.1109/JSAC.2008.081002}
}

@article{qin2018sparse,
  author  = {Zhijin Qin and
             Jianjun Fan and
             Yu Liu and
             Yue Gao and
             Geoffrey Ye Li},
  title   = {Sparse Representation for Wireless Communications: {A} Compressive Sensing Approach},
  journal = {{IEEE} Signal Process. Mag.},
  volume  = {35},
  number  = {3},
  pages   = {40--58},
  month   = may,
  year    = {2018},
  doi     = {10.1109/MSP.2018.2789521}
}

@article{gao2018compressive,
  author  = {Zhen Gao and
             Linglong Dai and
             Shuangfeng Han and
             Chih{-}Lin I and
             Zhaocheng Wang and
             Lajos Hanzo},
  title   = {Compressive Sensing Techniques for Next-Generation Wireless Communications},
  journal = {{IEEE} Wireless Commun.},
  volume  = {25},
  number  = {3},
  pages   = {144--153},
  month   = jun,
  year    = {2018},
  doi     = {10.1109/MWC.2017.1700147}
}

@ARTICLE{Joint_Sparse_and_Low_Rank,
  author={Li, Xingjian and Fang, Jun and Li, Hongbin and Wang, Pu},
  journal=IEEE_J_WCOM, 
  title={Millimeter Wave Channel Estimation via Exploiting Joint Sparse and Low-Rank Structures}, 
  year={2018},
  volume={17},
  number={2},
  pages={1123-1133},
  doi={10.1109/TWC.2017.2776108}
}

@ARTICLE{Low_rank_and_Sparse_Channel_Estimation,
  author={Gao, Zhen and Dai, Linglong and Wang, Zhaocheng and Chen, Sheng},
  journal=IEEE_J_SP, 
  title={Spatially Common Sparsity Based Adaptive Channel Estimation and Feedback for {FDD} Massive {MIMO}}, 
  year={2015},
  volume={63},
  number={23},
  pages={6169-6183},
  doi={10.1109/TSP.2015.2463260},
  }

@ARTICLE{zhou2017lowranktensor,
  author={Zhou, Zhou and Fang, Jun and Yang, Linxiao and Li, Hongbin and Chen, Zhi and Blum, Rick S.},
  journal={{IEEE} J. Sel. Areas Commun.},
  title={Low-Rank Tensor Decomposition-Aided Channel Estimation for Millimeter Wave {MIMO-OFDM} Systems},
  year={2017},
  volume={35},
  number={7},
  pages={1524--1538},
  doi={10.1109/JSAC.2017.2699338}}

@ARTICLE{shao2022exploiting,
  author={Shao, Xiaodan and Chen, Xiaoming and Zhong, Caijun and Zhang, Zhaoyang},
  journal={{IEEE} Trans. Wireless Commun.},
  title={Exploiting Simultaneous Low-Rank and Sparsity in Delay-Angular Domain for Millimeter-Wave/{Terahertz} Wideband Massive Access},
  year={2022},
  volume={21},
  number={4},
  pages={2336--2351},
  doi={10.1109/TWC.2021.3111225}}

@inproceedings{zhong2015efficient,
  title={Efficient matrix sensing using rank-1 gaussian measurements},
  author={Zhong, Kai and Jain, Prateek and Dhillon, Inderjit S},
  booktitle={Int. Conf. Algor. Learning Theory (ALT)},
  pages={3--18},
  year={2015},
  doi={10.1007/978-3-319-24486-0_1},
  
}

@InProceedings{pmlr-v89-li19e,
  title = 	 {Nonconvex Matrix Factorization from Rank-One Measurements},
  author =       {Li, Yuanxin and Ma, Cong and Chen, Yuxin and Chi, Yuejie},
  booktitle = 	 {Proc. Int.  Conf. Artificial Intelligence and Statistics},
  pages = 	 {1496--1505},
  year = 	 {2019},
  volume = 	 {89},
}

@INPROCEEDINGS{ms4l2o,
  author={Liu, Jialin and Chen, Xiaohan and Wang, Zhangyang and Yin, Wotao and Cai, Hanqin},
  booktitle={Proc. Int. Conf. Mach. Learn. (ICML)},
  title={Towards Constituting Mathematical Structures for Learning to Optimize},
  year={2023},
  pages={21426-21449},
  volume={202},
  month={Jul.}}

@ARTICLE{fista,
  author={Beck, Amir and Teboulle, Marc},
  journal={SIAM J. Imaging Sci.},
  title={A Fast Iterative Shrinkage-Thresholding Algorithm for Linear Inverse Problems},
  year={2009},
  volume={2},
  number={1},
  pages={183-202},
  publisher={SIAM},
  doi={10.1137/080716542}}

@ARTICLE{rank_one_prior,
  author={Liu, Jun and Liu, Ryan Wen and Sun, Jianing and Zeng, Tieyong},
  journal=IEEE_J_PAMI , 
  title={Rank-One Prior: Real-Time Scene Recovery}, 
  year={2023},
  volume={45},
  number={7},
  pages={8845-8860},
  doi={10.1109/TPAMI.2022.3226276}}

@ARTICLE{DCRNet,
  author={Tang, Shunpu and Xia, Junjuan and Fan, Lisheng and Lei, Xianfu and Xu, Wei and Nallanathan, Arumugam},
  journal={{IEEE} Trans. Veh. Technol.},
  title={Dilated Convolution Based {CSI} Feedback Compression for Massive {MIMO} Systems},
  year={2022},
  volume={71},
  number={10},
  pages={11216-11221},
  doi={10.1109/TVT.2022.3183596}}

@ARTICLE{cost2100,
  author={Liu, Lingfeng and Oestges, Claude and Poutanen, Juho and Haneda, Katsuyuki and Vainikainen, Pertti and Quitin, Fran{\c{c}}ois and Tufvesson, Fredrik and De Doncker, Philippe},
  journal={{IEEE} Wireless Commun.},
  title={The {COST} 2100 {MIMO} Channel Model},
  year={2012},
  volume={19},
  number={6},
  pages={92-99},
  doi={10.1109/MWC.2012.6393523}}

@ARTICLE{directional_training,
author={Zhang, Xing and Zhong, Lin and Sabharwal, Ashutosh},
journal={IEEE Trans. Wireless Commun.},
title={Directional Training for {FDD} Massive {MIMO}},
year={2018},
volume={17},
number={8},
pages={5183-5197},
doi={10.1109/TWC.2018.2838600}}

@ARTICLE{massive_mimo_channels,
author={Du, Xu and Sabharwal, Ashutosh},
  journal={IEEE Trans. Veh. Technol.},
  title={Massive {MIMO} Channels with Inter-User Angle Correlation: Open-Access Dataset, Analysis and Measurement-Based Validation},
  year={2022},
  volume={71},
  number={2},
  pages={1602--1616},
  doi={10.1109/TVT.2021.3131606}}

@ARTICLE{omp_ref,
  author={Tropp, Joel A. and Gilbert, Anna C.},
  journal={IEEE Trans. Inf. Theory},
  title={Signal Recovery from Random Measurements via Orthogonal Matching Pursuit},
  year={2007},
  volume={53},
  number={12},
  pages={4655-4666},
  doi={10.1109/TIT.2007.909108}}

@ARTICLE{wei2021randomized,
  author={Wei, Ziping and Li, Haozhan and Liu, Hongfu and Li, Bin and Zhao, Chenglin},
  journal={IEEE Commun. Lett.},
  title={Randomized Low-Rank Approximation Based Massive {MIMO} {CSI} Compression},
  year={2021},
  volume={25},
  number={6},
  pages={2004-2008},
  doi={10.1109/LCOMM.2021.3065751}}

@ARTICLE{CsiNet,
  author={Wen, Chao-Kai and Shih, Wan-Ting and Jin, Shi},
  journal={IEEE Wireless Commun. Lett.},
  title={Deep Learning for Massive {MIMO} {CSI} Feedback},
  year={2018},
  volume={7},
  number={5},
  pages={748-751},
  doi={10.1109/LWC.2018.2818160}}

@article{guo2022overview,
  author  = {Jiajia Guo and
             Chao{-}Kai Wen and
             Shi Jin and
             Geoffrey Ye Li},
  title   = {Overview of Deep Learning-Based {CSI} Feedback in Massive {MIMO} Systems},
  journal = {{IEEE} Trans. Commun.},
  volume  = {70},
  number  = {12},
  pages   = {8017--8045},
  month   = dec,
  year    = {2022},
  doi     = {10.1109/TCOMM.2022.3217777}
}

@article{he2019modeldriven,
  author  = {Hengtao He and
             Shi Jin and
             Chao{-}Kai Wen and
             Feifei Gao and
             Geoffrey Ye Li and
             Zongben Xu},
  title   = {Model-Driven Deep Learning for Physical Layer Communications},
  journal = {{IEEE} Wireless Commun.},
  volume  = {26},
  number  = {5},
  pages   = {77--83},
  year    = {2019},
  doi     = {10.1109/MWC.2019.1800447}
}

@article{cao2023deepunfolding,
  author  = {Zhenyu Cao and
             Jiajia Guo and
             Chao{-}Kai Wen and
             Shi Jin},
  title   = {Deep-Unfolding-Based Bit-Level {CSI} Feedback in Massive {MIMO} Systems},
  journal = {{IEEE} Wireless Commun. Lett.},
  volume  = {12},
  number  = {2},
  pages   = {371--375},
  year    = {2023},
  doi     = {10.1109/LWC.2022.3227315}
}

@INPROCEEDINGS{CRNet,
  author={Lu, Zhilin and Wang, Jintao and Song, Jian},
  booktitle={Proc. {IEEE} Int. Conf. Commun. ({ICC})},
  title={Multi-Resolution {CSI} Feedback With Deep Learning in Massive {MIMO} System},
  year={2020},
  pages={1-6},
  doi={10.1109/ICC40277.2020.9149229}}

@ARTICLE{CLNet,
  author={Ji, Sijie and Li, Mo},
  journal={IEEE Wireless Commun. Lett.},
  title={{CLNet}: Complex Input Lightweight Neural Network Designed for Massive {MIMO} {CSI} Feedback},
  year={2021},
  volume={10},
  number={10},
  pages={2318-2322},
  doi={10.1109/LWC.2021.3100493}}

@ARTICLE{ACRNet,
  author={Lu, Zhilin and Zhang, Xudong and He, Hongyi and Wang, Jintao and Song, Jian},
  journal={IEEE Trans. Wireless Commun.},
  title={Binarized Aggregated Network With Quantization: Flexible Deep Learning Deployment for {CSI} Feedback in Massive {MIMO} Systems},
  year={2022},
  volume={21},
  number={7},
  pages={5514-5525},
  doi={10.1109/TWC.2022.3141653}}

@ARTICLE{DSNLCsiNet,
  author={Yu, Xiaotong and Li, Xiangyi and Wu, Huaming and Bai, Yang},
  journal={IEEE Commun. Lett.},
  title={{DS-NLCsiNet}: Exploiting Non-Local Neural Networks for Massive {MIMO} {CSI} Feedback},
  year={2020},
  volume={24},
  number={12},
  pages={2790-2794},
  doi={10.1109/LCOMM.2020.3019653}}

@ARTICLE{CsiNetPlus,
  author={Guo, Jiajia and Wen, Chao-Kai and Jin, Shi and Li, Geoffrey Ye},
  journal={IEEE Trans. Wireless Commun.},
  title={Convolutional Neural Network-Based Multiple-Rate Compressive Sensing for Massive {MIMO} {CSI} Feedback: Design, Simulation, and Analysis},
  year={2020},
  volume={19},
  number={4},
  pages={2827-2840},
  doi={10.1109/TWC.2020.2968430}}

@ARTICLE{TransNet,
  author={Cui, Yaodong and Guo, Aihuang and Song, Chunlin},
  journal={IEEE Wireless Commun. Lett.},
  title={{TransNet}: Full Attention Network for {CSI} Feedback in {FDD} Massive {MIMO} System},
  year={2022},
  volume={11},
  number={5},
  pages={903-907},
  doi={10.1109/LWC.2022.3149416}}

@ARTICLE{SSE-Net,
  author={Liu, Mingyi and Zhang, Aihua and Shi, Wensen and Yao, Dan and Yang, Shuaishuai},
  journal={IEEE Wireless Commun. Lett.},
  title={Structured Spatial Encoding Network for Efficient {CSI} Feedback in {FDD} Massive {MIMO} Systems},
  year={2026},
  volume={15},
  pages={2219-2223},
  doi={10.1109/LWC.2026.3675355}}

@MISC{TCLNet,
  author={Yang, Zijiu and Yang, Qianqian and Tang, Shunpu and Yang, Tingting and Shi, Zhiguo},
  title={{TCLNet}: A Hybrid Transformer--{CNN} Framework Leveraging Language Models as Lossless Compressors for {CSI} Feedback},
  howpublished={arXiv:2601.06588},
  year={2026},
  }

@article{LLMCsiNet,
  author={Wu, Jie and Xu, Wei and Liang, Le and You, Xiaohu and Debbah, M{\'e}rouane},
  title={Large Language Model Empowered {CSI} Feedback in Massive {MIMO} Systems},
  journal={{IEEE} Trans. Wireless Commun.},
  volume={25},
  pages={13835--13851},
  year={2026},
  doi={10.1109/TWC.2026.3671802},
  note={arXiv:2603.02686}}

@article{cai2015rop,
  title={ROP: Matrix Recovery via Rank-One Projections},
  author={Cai, T. Tony and Zhang, Anru},
  journal={The Annals of Statistics},
  volume={43},
  number={1},
  pages={102--138},
  year={2015},
  doi={10.1214/14-AOS1267}
}

@InProceedings{hu2018squeeze,
  author={Hu, Jie and Shen, Li and Sun, Gang},
  title={Squeeze-and-Excitation Networks},
  booktitle={Proc. IEEE Conf. Comput. Vis. Pattern Recognit. (CVPR)},
  pages={7132--7141},
  month={Jun.},
  year={2018},
  doi={10.1109/CVPR.2018.00745}
}

@inproceedings{ronneberger2015unet,
  title     = {U-Net: Convolutional Networks for Biomedical Image Segmentation},
  author    = {Ronneberger, Olaf and Fischer, Philipp and Brox, Thomas},
  booktitle = {Medical Image Computing and Computer-Assisted Intervention (MICCAI)},
  series    = {Lecture Notes in Computer Science},
  volume    = {9351},
  pages     = {234--241},
  publisher = {Springer},
  year      = {2015},
  doi       = {10.1007/978-3-319-24574-4_28},
}

@ARTICLE{Cao2021Lightweight,
  author={Cao, Zheng and Shih, Wan-Ting and Guo, Jiajia and Wen, Chao-Kai and Jin, Shi},
  journal={IEEE Commun. Lett.}, 
  title={Lightweight Convolutional Neural Networks for {CSI} Feedback in Massive {MIMO}}, 
  year={2021},
  volume={25},
  number={8},
  pages={2624-2628},
  doi={10.1109/LCOMM.2021.3076504}}

@ARTICLE{Sun2021Lightweight,
  author={Sun, Yuyao and Xu, Wei and Liang, Le and Wang, Ning and Li, Geoffrey Ye and You, Xiaohu},
  journal={IEEE Wireless Communications Letters}, 
  title={A Lightweight Deep Network for Efficient {CSI} Feedback in Massive {MIMO} Systems}, 
  year={2021},
  volume={10},
  number={8},
  pages={1840-1844},
  doi={10.1109/LWC.2021.3083331}}

@INPROCEEDINGS{Cheng2024SwinCFNet,
  author        = {Jiaming Cheng and Wei Chen and Jialong Xu and Yiran Guo and Lun Li and Bo Ai},
  title         = {Swin Transformer-Based {CSI} Feedback for Massive {MIMO}},
  year          = {2023},
  booktitle     = {Proc. IEEE Int. Conf. Commun. Technol. (ICCT)},
  pages         = {809--814},
  doi           = {10.1109/ICCT59356.2023.10419637},
  note          = {arXiv:2401.06435},
}

@ARTICLE{WiFoCF,
  author={Liu, Xuanyu and Gao, Shijian and Liu, Boxun and Cheng, Xiang and Yang, Liuqing},
  journal={IEEE Trans. Wireless Commun.}, 
  title={WiFo-CF: Wireless Foundation Model for CSI Feedback}, 
  year={2026},
  volume={25},
  pages={15039-15053},
  doi={10.1109/TWC.2026.3681311}}

@article{Guo2022ModelDriven,
  author        = {Jianhua Guo and Lei Wang and Feng Li and Jiang Xue},
  title         = {{CSI} Feedback With Model-Driven Deep Learning of Massive {MIMO} Systems},
  journal       = {IEEE Commun. Lett.},
  year          = {2022},
  volume        = {26},
  number        = {3},
  pages         = {547--551},
  doi           = {10.1109/LCOMM.2021.3138927},
}

@INPROCEEDINGS{Zhang2023ModelDriven,
  author={Zhang, Yangyang and Zhang, Xichang and Liu, Yi},
  booktitle={IEEE Int. Conf. Commun. (ICC)}, 
  title={Model-Driven Lightweight Network for CSI Feedback in Time-Varying Massive MIMO Systems}, 
  year={2023},
  pages={4961-4966},
  doi={10.1109/ICC45041.2023.10278973}
}

@ARTICLE{9768327,
  author={Yin, Ziqing and Xu, Wei and Xie, Renjie and Zhang, Shaoqing and Ng, Derrick Wing Kwan and You, Xiaohu},
  journal={IEEE Trans. Wireless Commun.}, 
  title={Deep CSI Compression for Massive MIMO: A Self-Information Model-Driven Neural Network}, 
  year={2022},
  volume={21},
  number={10},
  pages={8872-8886},
  doi={10.1109/TWC.2022.3170576}}

@ARTICLE{Chen2025CSICompression,
  author={Chen, Jie and Hillery, William J. and Mestav, Kursat Rasim and Nuzman, Carl and Saniee, Iraj and Xing, Yunchou},
  journal={IEEE Wireless Commun.}, 
  title={CSI Compression for Massive MIMO: Model-Based or Data-Driven?}, 
  year={2025},
  volume={32},
  number={1},
  pages={22-27},
  doi={10.1109/MWC.010.2400078}}

@ARTICLE{9513579,
  author={Liu, Zhenyu and del Rosario, Mason and Ding, Zhi},
  journal={IEEE Trans. Wireless Commun.}, 
  title={A Markovian Model-Driven Deep Learning Framework for Massive MIMO CSI Feedback}, 
  year={2022},
  volume={21},
  number={2},
  pages={1214-1228},
  doi={10.1109/TWC.2021.3103120}}

@INPROCEEDINGS{Cui2025LLM,
  author={Cui, Yiming and Guo, Jiajia and Wen, Chao-Kai and Jin, Shi and Tong, En},
  booktitle={IEEE Global Commun. Conf. (GLOBECOM)}, 
  title={Exploring the Potential of Large Language Models for Massive MIMO CSI Feedback}, 
  year={2025},
  pages={3511-3516},
  doi={10.1109/GLOBECOM59602.2025.11432525}}

@ARTICLE{Guo2026PromptEnabled,
  author={Guo, Jiajia and Cui, Yiming and Wen, Chao-Kai and Jin, Shi},
  journal={IEEE J. Sel. Areas Commun.}, 
  title={Prompt-Enabled Large AI Models for CSI Feedback}, 
  year={2026},
  volume={44},
  pages={2654-2668},
  doi={10.1109/JSAC.2025.3643823}}

@ARTICLE{Ma2025LowComplexity,
  author={Ma, Yifan and He, Hengtao and Song, Shenghui and Zhang, Jun and Letaief, Khaled B.},
  journal={IEEE Trans. Wireless Commun.}, 
  title={Low-Complexity CSI Feedback for FDD Massive MIMO Systems via Learning to Optimize}, 
  year={2025},
  volume={24},
  number={4},
  pages={3483-3498},
  doi={10.1109/TWC.2025.3531447}}

@ARTICLE{Vlachos2019Wideband,
  author={Vlachos, Evangelos and Alexandropoulos, George C. and Thompson, John},
  journal={IEEE J. Sel. Topics Signal Process.}, 
  title={Wideband MIMO Channel Estimation for Hybrid Beamforming Millimeter Wave Systems via Random Spatial Sampling}, 
  year={2019},
  volume={13},
  number={5},
  pages={1136-1150},
  doi={10.1109/JSTSP.2019.2937633}}

@ARTICLE{Shen2015Joint,
  author={Shen, Wenqian and Dai, Linglong and Shim, Byonghyo and Mumtaz, Shahid and Wang, Zhaocheng},
  journal={IEEE Commun. Lett.}, 
  title={Joint CSIT Acquisition Based on Low-Rank Matrix Completion for FDD Massive MIMO Systems}, 
  year={2015},
  volume={19},
  number={12},
  pages={2178-2181},
  doi={10.1109/LCOMM.2015.2492960}}

@article{Cui2024NearFieldBeamforming,
  author  = {Mingyao Cui and Linglong Dai},
  title   = {Near-Field Wideband Beamforming for
             Extremely Large Antenna Arrays},
  journal = {IEEE Trans. Wireless Commun.},
  volume  = {23},
  number  = {10},
  pages   = {13110--13124},
  year    = {2024},
}

% =========================
% Supplementary Material
% =========================

\clearpage

\setcounter{page}{1}
\setcounter{section}{0}

\twocolumn[
\begin{@twocolumnfalse}

\begin{center}
    {\LARGE \bfseries
    Supplementary Material for\\[3pt]
    ``Model-Driven Deep Learning with Rank-One Sensing for Efficient CSI Feedback''
    \par}

    \vspace{1.2em}

    {\large
    Shunpu Tang,
Qianqian Yang,
Seung-Woo Ko,
Jihong Park,
and Kaibin Huang
    \par}
\end{center}

\vspace{1.5em}

\end{@twocolumnfalse}
]

\section{RT-CSI Dataset Configuration}\label{app:rt_scenarios}

  The RT-CSI benchmark is generated with NVIDIA Sionna RT~2.0.1. To construct a mixed-scene dataset for training and evaluation, we use five non-Shenzhen scenes, including Munich, Paris \'Etoile, Florence, ZJU, and SUTD. The Munich, Paris \'Etoile, and Florence layouts are the OpenStreetMap (OSM)-derived scenes released with Sionna RT, whereas the ZJU, SUTD, and Shenzhen Futian CBD layouts are constructed from OSM exports using the Blosm Blender add-on. The Shenzhen Futian CBD scene is kept as a held-out test scenario to evaluate the generalization capability of CSI feedback models under dense high-rise deployment, where strong blockage and diffuse multipath are expected. Unless otherwise specified, all scenes follow the common carrier, antenna, propagation, and normalization settings summarized in Table~\ref{tab:rt_csi_params}.
  
  For BS deployment, we place $N_{\text{BS}}=5$ BSs in each non-Shenzhen scene by selecting rooftop candidates ranked by height, while enforcing a minimum lateral separation between any two BSs to avoid clustering co-located transmitters. For the ZJU and SUTD campus scenes, the rooftop candidate set is further restricted to the campus bounding box. This restriction excludes nearby residential blocks in ZJU and the Singapore Expo exhibition halls adjacent to SUTD, and selects the central campus library as one of the five BSs in both campus scenes. In contrast, Shenzhen Futian CBD uses a separate held-out configuration with $N_{\text{BS}}=10$ BSs. The BS heights are constrained to $25$--$50~\mathrm{m}$ so that supertall landmarks, including the Ping An Finance Center and KK100, act as dominant blockers and scatterers rather than transmitter sites.

\begin{table}[!t]
  \centering
  \caption{Common RT-CSI generation parameters.}
  \label{tab:rt_csi_params}
  \footnotesize
  \setlength{\tabcolsep}{3pt}
  \renewcommand{\arraystretch}{1.05}
  \begin{threeparttable}
  \begin{tabularx}{\columnwidth}{@{}lX@{}}
  \toprule
  \textbf{Parameter} & \textbf{Configuration} \\
  \midrule
  Engine & Sionna RT 2.0.1 \\
  Carrier frequency & 3.5 GHz \\
  Bandwidth & 20 MHz \\
  Subcarriers & 1024 \\
  BS array & $1 \times 32$ ULA, vertical polarization, isotropic pattern \\
  UE array & Single antenna, vertical polarization, isotropic pattern \\
  Antenna spacing & $0.5\lambda$ \\
  BSs per scene & 5 for non-Shenzhen scenes; 10 for Shenzhen \\
  UE height & $1.5~\mathrm{m}$ above local terrain \\
  RX sampling grid & $1~\mathrm{m}\times1~\mathrm{m}$ \\
  PathSolver maximum depth & 5 for non-Shenzhen scenes; 7 for Shenzhen \\
  Propagation effects & LoS, specular reflection, refraction, diffraction \\
  Diffuse reflection & Disabled for non-Shenzhen scenes; enabled for Shenzhen\tnote{b} \\
  Synthetic array & Enabled \\
  Normalization & Per-sample max-abs scaling of real/imaginary angular-delay coefficients to $[0,1]$ \\
  \bottomrule
  \end{tabularx}
  \begin{tablenotes}\footnotesize
  \item[b] For Shenzhen Futian CBD, diffuse reflection is enabled with Lambertian scattering coefficient $\sigma_{\mathrm{diff}}
  =0.1$ applied uniformly to all ITU materials.
  \end{tablenotes}
  \end{threeparttable}
  \end{table}

  After the BS locations are determined, UE candidates are sampled on a $1~\mathrm{m}\times1~\mathrm{m}$ grid over the union of disks centered at the BS locations. Each UE is placed at $z=z_{\mathrm{terrain}}+1.5~\mathrm{m}$, where $z_{\mathrm{terrain}}$ denotes the local terrain elevation. For the five non-Shenzhen scenes, the aggregate gain from BS $b$ is computed as $g_b=\sum_{n_t,n_c}|\mathbf{H}_{b,n_t,n_c}|^2$ over all transmit antennas and subcarriers. The serving BS is then selected by $b^\star=\arg\max_b g_b$, and only the wideband channel from $b^\star$ is stored. For Shenzhen, we instead store all valid UE-BS links and tag each sample with its BS index, which supports cross-BS and few-shot adaptation analyses. Before storage, each angular-delay CSI tensor is normalized per sample by the maximum absolute real or imaginary coefficient and shifted to $[0,1]$.

The held-out Shenzhen Futian CBD scene is configured as a dense high-rise stress case. In addition to the BS-height constraint, diffuse reflection is enabled for all ITU materials with $\sigma_{\mathrm{diff}}=0.1$, while the default Sionna RT setting uses $\sigma_{\mathrm{diff}}=0$. We set $R_{\mathrm{UE}}=400~\mathrm{m}$ around each BS and increase the PathSolver depth to 7 to capture richer multipath propagation. The fixed Shenzhen test set contains $40~\mathrm{k}$ valid UE-BS channel samples and is used exclusively for testing. For few-shot adaptation, separate Shenzhen training sets are generated and split into training and validation subsets with an $80/20$ ratio.

\section{Visualizations}\label{app:lrp_visualization}

\begin{figure*}[!t]
    \centering
    \includegraphics[width=\textwidth]{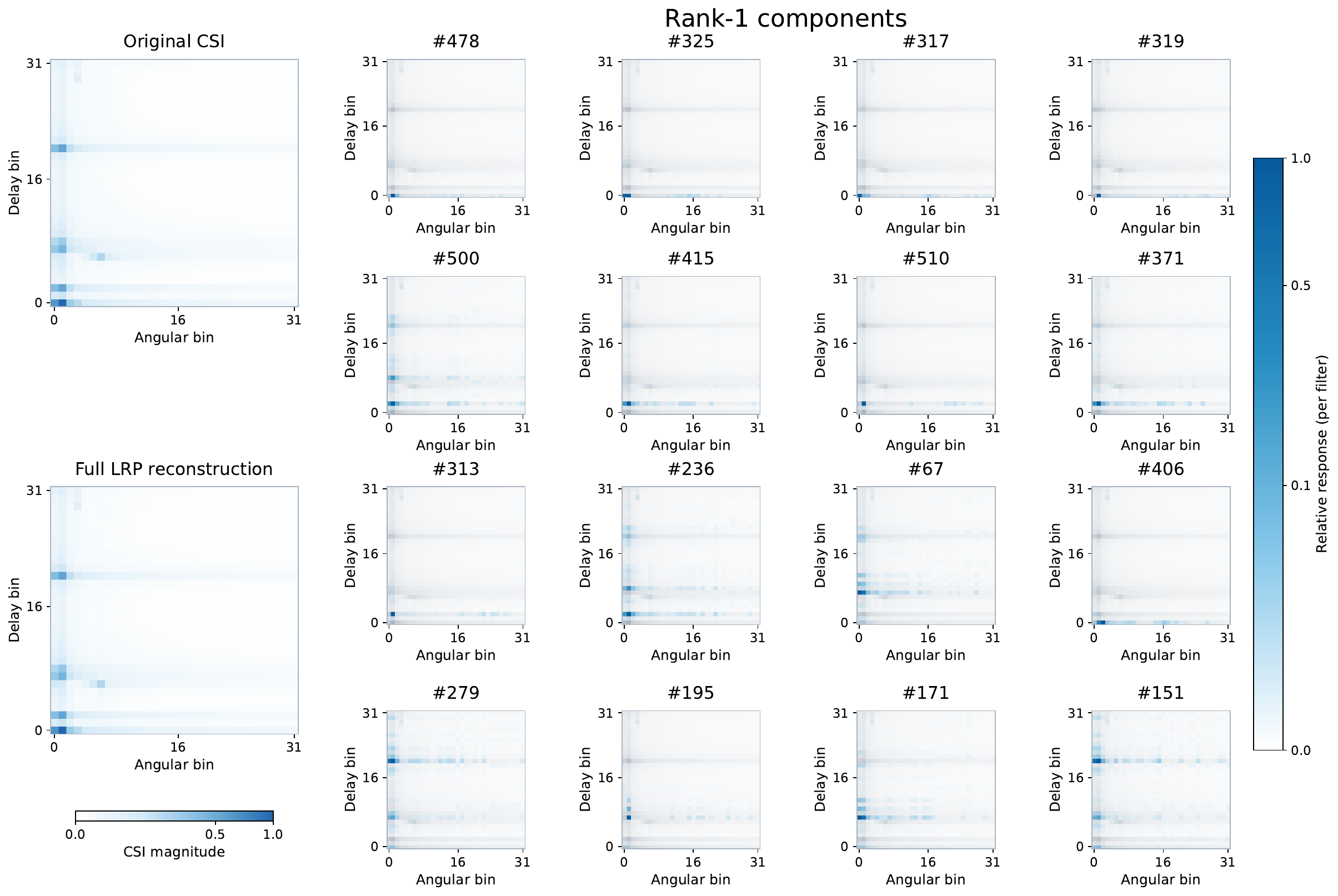}
    \caption{Original angular-delay CSI, full LRP reconstruction, and 16 selected rank-one responses. The numbers above the response maps denote filter indices. CSI magnitude and relative response per filter use separate color scales.}
    \label{fig:lrp_sample_match}
\end{figure*}

To further analyze how LRP captures the channel structure, we visualize an angular-delay CSI sample, its full LRP reconstruction, and 16 selected rank-one responses in Fig.~\ref{fig:lrp_sample_match}. From this figure, we can see that the full reconstruction preserves the main delay-domain energy concentrations and their angular distribution. The selected responses exhibit different delay--angular patterns, with several responses concentrated around the prominent delay bins of the original CSI. This observation is consistent with the matched-filter interpretation in \eqref{eq:lpa_rank_one_sensing}, where each sensing branch measures the alignment between the channel and a learnable separable pattern. These results illustrate how the proposed LRP architecture captures structured channel features through rank-one sensing.

\begin{figure*}[!t]
    \centering
    \includegraphics[width=\textwidth]{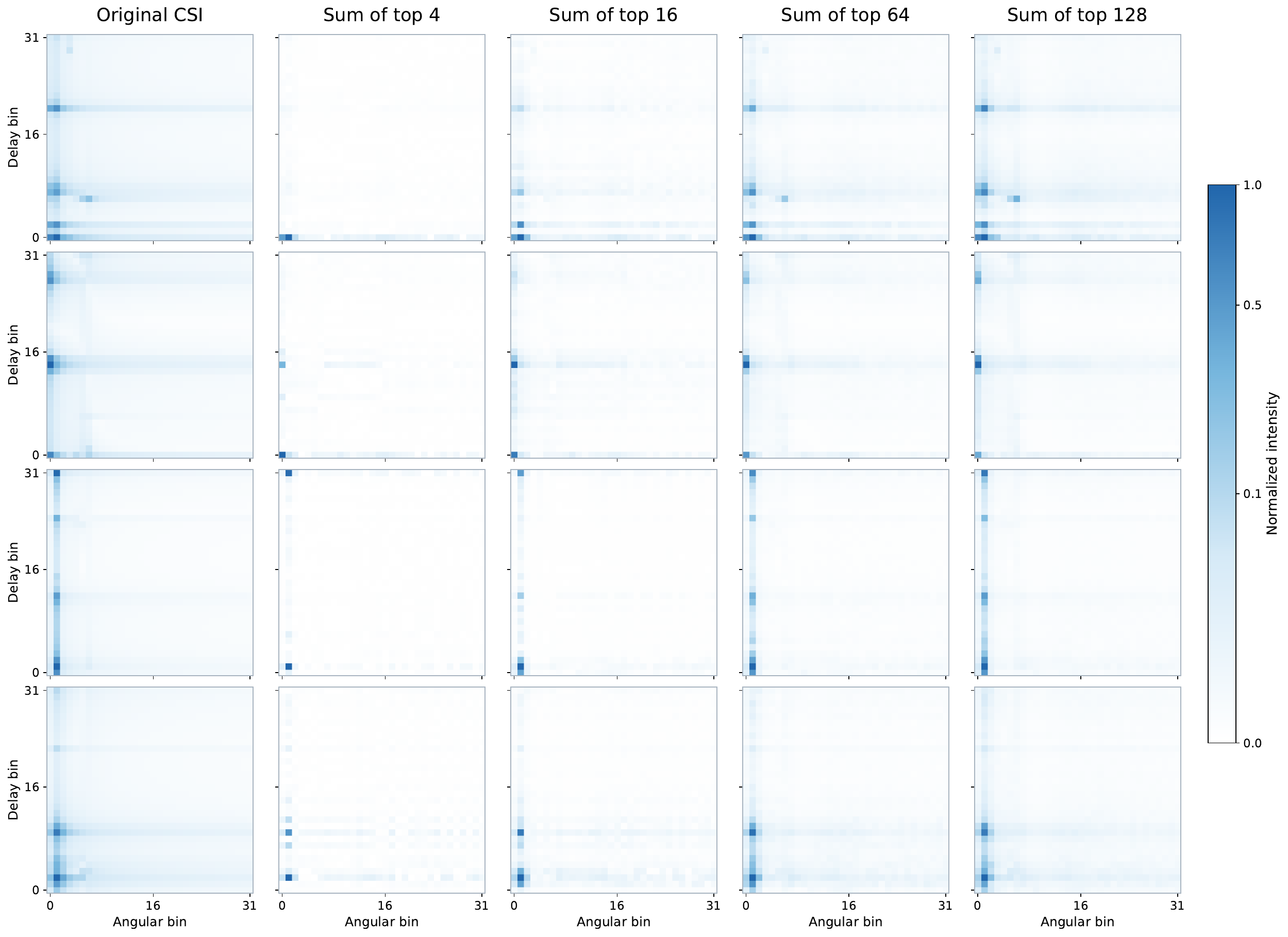}
    \caption{Original angular-delay CSI and the sums of the top 4, 16, 64, and 128 LRP responses for four samples, one per row. The color scale indicates normalized intensity.}
    \label{fig:lrp_sample_match_sum}
\end{figure*}

Fig.~\ref{fig:lrp_sample_match_sum} further shows the sums of the top 4, 16, 64, and 128 LRP responses for four CSI samples with different delay--angular energy distributions. From this figure, we can see that the sum of the top 4 responses captures only part of the channel structure, while including more responses recovers additional energy concentrations and weaker details. For example, in the first sample, the sums of the top 64 and 128 responses reproduce the main horizontal delay-domain clusters more closely than the sums of the top 4 and 16 responses. A similar trend can be observed in the other samples. These results show that multiple rank-one responses contribute complementary channel information, supporting the use of a bank of learnable sensing patterns in the LRP encoder.

\begin{figure*}[!t]
    \centering
    \includegraphics[width=\textwidth]{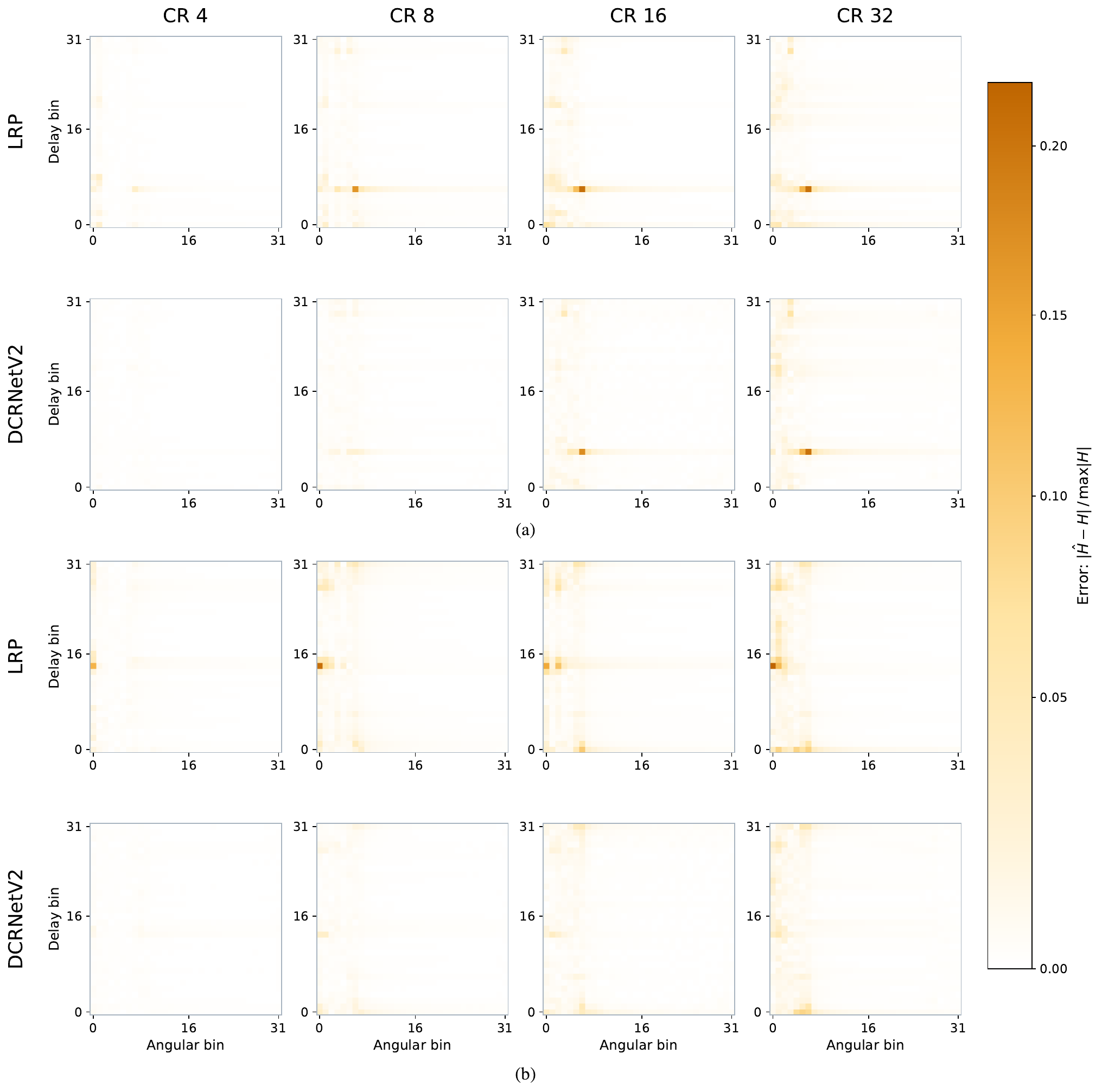}
    \caption{Normalized absolute reconstruction errors of LRP and DCRNetV2 for two CSI samples in (a) and (b) at \(\eta\in\{1/4,1/8,1/16,1/32\}\). The column labels use \(\mathrm{CR}=1/\eta\). Errors are normalized by the maximum CSI magnitude of each sample, with a common color scale across all panels.}
    \label{fig:lrp_dcrnet_sample_errors}
\end{figure*}

In Fig.~\ref{fig:lrp_dcrnet_sample_errors}, we further compare the reconstruction errors of LRP and DCRNetV2 for two CSI samples at \(\eta\in\{1/4,1/8,1/16,1/32\}\). Errors are normalized by the maximum CSI magnitude of each sample, with a common color scale. From this figure, we can see that DCRNetV2 reduces the localized errors remaining in the LRP reconstruction. Specifically, for sample (a), the error concentration at small delay bins is weaker with DCRNetV2, particularly at \(\eta=1/8\). For sample (b), DCRNetV2 reduces the prominent error concentration around the middle delay bins across the tested compression ratios. As \(\eta\) decreases, residual errors remain visible for both methods. These observations are consistent with the ablation results in Table~\ref{tab:ablation}, and support the design of DCRNetV2, where the LRP backbone captures the dominant channel structure while the residual branches compensate for the remaining details.

\end{document}